\documentclass{emulateapj}

\newcommand{\km}{${\rm km\,s}^{-1}$}
\newcommand{\fuse}{{\em FUSE}}
\newcommand{\hi}{H$\;${\small\rm I}\relax}
\newcommand{\hii}{H$\;${\small\rm II}\relax}

\newcommand{\ari}{Ar$\;${\small\rm I}\relax}

\newcommand{\cii}{C$\;${\small\rm II}\relax}

\newcommand{\ciii}{C$\;${\small\rm III}\relax}
\newcommand{\civ}{C$\;${\small\rm IV}\relax}
\newcommand{\nni}{N$\;${\small\rm I}\relax}
\newcommand{\nii}{N$\;${\small\rm II}\relax}
\newcommand{\niii}{N$\;${\small\rm III}\relax}

\newcommand{\oi}{O$\;${\small\rm I}\relax}

\newcommand{\ovi}{O$\;${\small\rm VI}\relax}

\newcommand{\si}{S$\;${\small\rm I}\relax}
\newcommand{\sii}{S$\;${\small\rm II}\relax}
\newcommand{\siii}{Si$\;${\small\rm II}\relax}
\newcommand{\siiii}{Si$\;${\small\rm III}\relax}
\newcommand{\Siii}{S$\;${\small\rm III}\relax}
\newcommand{\siiv}{Si$\;${\small\rm IV}\relax}

\newcommand{\feii}{Fe$\;${\small\rm II}\relax}
\newcommand{\feiii}{Fe$\;${\small\rm III}\relax}

\newcommand{\lya}{Ly\,$\alpha$\relax}
\newcommand{\lyb}{Ly\,$\beta$\relax}

\slugcomment{Accepted for publication in the ApJ}
\shortauthors{Lehner et al.}
\shorttitle{Metallicity and Ionization of the Bridge}
\begin{document}

\title{Metallicity and Physical Conditions in the Magellanic Bridge\altaffilmark{1} }
\author{N.\ Lehner\altaffilmark{2},
	J.C. \ Howk\altaffilmark{2},
	F.P. \ Keenan\altaffilmark{3},
	J.V. Smoker\altaffilmark{3}
	}
   
\altaffiltext{1}{Based on observations made with the NASA-CNES-CSA 
Far Ultraviolet Spectroscopic Explorer. FUSE is operated for NASA by the Johns 
Hopkins University under NASA contract NAS5-32985. Based on observations made with the NASA/ESA Hubble Space Telescope,
obtained at the Space Telescope Science Institute, which is operated by the
Association of Universities for Research in Astronomy, Inc. under NASA
contract No. NAS5-26555.}
\altaffiltext{2}{Department of Physics, University of Notre Dame, 225 Nieuwland Science Hall, Notre Dame, IN 46556}
\altaffiltext{3}{Astrophysics Research Centre, School of Mathematics and Physics, The Queen's University of Belfast, Belfast, UK.}

\begin{abstract}
We present a new analysis of the diffuse gas in the Magellanic Bridge (RA\,$\ga 3^{\rm h}$) based 
on {\em HST}/STIS E140M and {\fuse}\ spectra of 2 early-type stars lying within the Bridge 
and a QSO behind it. We derive the column densities of \hi\ (from \lya), 
\nni, \oi, \ari, \siii, \sii, and \feii\ of the gas in the Bridge. 
Using the atomic species, we determine the first gas-phase metallicity of the Magellanic Bridge,
$[{\rm Z/H}] = -1.02 \pm 0.07$ toward one sightline, and $-1.7 < [{\rm Z/H}] < -0.9$
toward the other one, a factor 2 or more smaller than the present-day SMC metallicity. 
Using the metallicity and $N($\hi$)$, 
we show that the Bridge gas along our three lines of sight is $\sim$70--90\% ionized, 
despite high \hi\ columns, $\log N($\hi$) \simeq 19.6$--20.1. 
Possible sources for the ongoing ionization are certainly the hot stars 
within the Bridge, hot gas (revealed by \ovi\ absorption), and leaking
photons from the SMC and LMC.
From the analysis of \cii*, we deduce that 
the overall density of the Bridge must be low ($<0.03$--0.1 cm$^{-3}$).
We argue that our findings combined with other recent observational results 
should motivate new models of the evolution of the SMC-LMC-Galaxy system.
\end{abstract}


\section{Introduction}
The evolution of galaxies is closely coupled to 
the interactions between them. Encounters
between galaxies are frequent and were certainly 
habitual in the early epoch of galaxy formation \citep[e.g.,][]{barnes92,larson78,maller06}. These 
encounters reshape the galaxies and transfer mass, energy, 
metals between the galaxies, and may even create new sites of star
formation within the newly produced gaseous features between the
galaxies or new burst of star-formations in the colliding
galaxies \citep{larson78,barton07,demello07}. These galactic interactions
are not believed to be a major contributor to 
the pollution of the intergalactic medium \citep[e.g.,][]{aguirre01}, 
but they nonetheless affect the metal content and physics of the galaxies 
themselves and their halos, and therefore the evolution of the galaxies
and their surroundings. 

In our galactic neighborhood, interactions
between the Magellanic clouds and the Galaxy are believed to have 
produced several large gaseous features: The Magellanic Bridge linking the Small
Magellanic Cloud (SMC) and the Large Magellanic Cloud (LMC), 
the Magellanic Stream and the Leading Arm apparenly linking the Clouds
and our Galaxy \citep[e.g.,][]{mathewson74,putman00,bruns05}.  
The Magellanic Bridge is believed to have been produced
through an interaction between the SMC and LMC, while the
Stream and Leading arm may have arisen through an interaction
between the Galaxy and the Clouds \citep[e.g.][]{gardiner96}, although 
the origin of these latter gaseous features is far from settled
\citep{nidever07,besla07}. The proximity of these gaseous features 
and the global view of the Magellanic System with little line-of-sight confusion
allow us to study the result of interacting galaxies in a way not 
otherwise possible. The Magellanic Bridge (hereafter Bridge) is the focus of 
the present work. 

In the past few years, multi-wavelength observations of the Bridge have 
revealed some key characteristics. Radio \hi\ observations have 
shown that approximately 2/3 of the \ion{H}{1} gas surrounding the Clouds is
found in the Magellanic Bridge, while only 25\% and 6\% are found in
the Stream and the Leading Arm, respectively \citep{bruns05}.  Some of the
Bridge gas appears even to build-up and feed the Stream via an Interface
Region \citep{bruns05}. Far-ultraviolet  observations have 
shown multiple gas phases in the Bridge including not only the well
known neutral gas, but also a significant amount of ionized gas and a small
H$_2$ content \citep[][this paper]{lehner01,lehner02}. Dense molecular 
clouds were also subsequently identified in form of CO \citep{muller03,mizuno06} 
(although these CO surveys were realized in the SMC Wing, a much denser region
in stars and gas content of the Bridge).  
Optical studies have revealed  massive hot stars throughout the Bridge \citep{demers98}. 
The ages of these stars are $\sim$10 to 40 Myr, implying that star formation 
is ongoing within the Bridge gas since these stars could not migrate from the SMC 
during their lifetimes, although \citet{harris07} suggests that star formation 
in the Bridge may  have been more important 200--300 Myr ago. 

The N-body numerical  simulation of the Galaxy-SMC-LMC system by
\citet{gardiner96} have reproduced some 
of the observed characterics of the Bridge (principally the \hi\ column density 
and velocity distributions), along with those of the Stream.
In this paper, we still use their numerical simulation as a basis, although
we note that the hypotheses and validity of this model and other recent numerical simulations 
\citep[e.g.,][]{yoshizawa03} has been put recently in question in view of the 
new {\em Hubble Space Telescope (HST)}\ measurements of the proper motions of the LMC
and SMC \citep[][and see \S\ref{sec-per}]{kallivayalil06,kallivayalil06b,besla07}. 
Gardiner \& Noguchi's model  predicts
that the Bridge was formed from tidally-stripped gas pulled from the SMC 
during a close encounter between the SMC and the LMC some 200 Myr ago. 
The new calculations of the LMC and SMC orbits still suggest that the closest 
approach between these two galaxies occured some 200 Myr ago \citep{kallivayalil06},
and therefore the new proper motions may only affect the interpretation of the 
origins of the Stream and Leading Arm. Nevertheless, there appear to be some
challenges to these models. 

Tidal models predict that both 
gas and stars are pulled from the SMC, and yet search for an old stellar population in 
the Bridge has so far failed \citep{harris07}. Furthermore, material pulled from the SMC should have a somewhat similar 
metallicity that the SMC some 200 Myr ago. But this appears in conflict with the B-type star 
abundances in the Bridge that imply an extremely low Bridge metallicity, $-1.1$ dex 
from solar \citep{rolleston99,lee05}, a factor of 3 and 5 metal deficiency with respect to the SMC and LMC, 
respectively. The metallicity in the sparse regions of the 
Bridge is also at odds with those measured in the SMC wing, also known as the 
Western end of the Bridge, where the metallicity of B-type stars is found to be 
SMC-like \citep{lee05}. These metallicity estimates complicate the origin of the Bridge and 
suggest different evolutions or origins for the sparse and dense stellar regions
of the Bridge. 

An estimate of the present-day metallicy of the Bridge that is independent of stellar measurements 
is therefore critical. In this work, we present an estimate of the absolute 
gas-phase abundances of the Bridge toward two stars situated in relatively low \hi\ column
density regions. We compare the column densities of the neutral 
species (\nni, \oi, \ari) observed in absorption in the Space Telescope Imaging Spectrograph (STIS)
and {\em Far Ultraviolet Spectroscopic Explorer} ({\fuse}) spectra to those for
\hi\ Ly$\alpha$ lines toward two early-type stars, i.e. our metallicity estimate is mostly
independent of ionization correction.  The Bridge Ly$\alpha$ absorption is heavily blended
with the Galactic component, but we show that it is feasible to 
retrieve the Bridge \hi\ column density. Using the metallicity and the amount
of \hi\ along these sightlines, we can for the first time quantify the amount
of ionized gas in the Bridge; ionized gas is revealed to be a major component of the Bridge,
yet this gas-phase is mostly ignored in numerical simulations.

Our paper is organized as follows. In \S\ref{sec-obs} we briefly discuss the 
previous STIS and {\fuse}\ observations, and the new {\fuse}\ observations of 
DGIK\,975 and DI\,1388. In \S\ref{sec-analysis} we discuss our measurements
of the metals and the \hi\ colum densities. We present in \S\ref{sec-result}
the first metallicity estimate of the Bridge gas and the fraction of neutral, ionized, and molecular
gas. From the \cii*\ diagnostic, we estimate the electron density and
cooling rate in the Bridge. In \S\ref{sec-discuss} we discuss our results (in particular 
address the possible origins for the low metallicity and ionization
sources of Bridge) and the current observational challenges to numerical simulations. 
Finally, in \S\ref{sec-sum} we summarize our key findings.

\section{Observations}\label{sec-obs}
\begin{figure*}[tbp]
\epsscale{0.7} 
\plotone{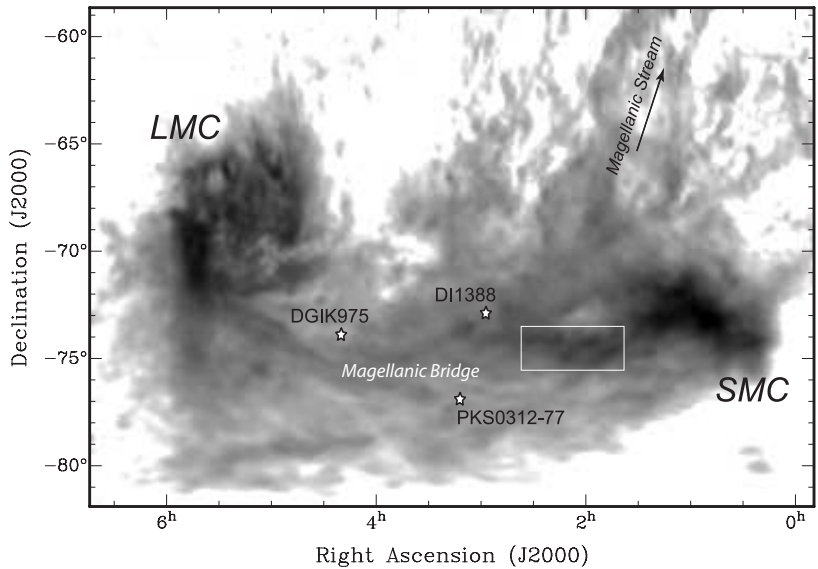}
\caption{Location of our targets superimposed on the \hi\ column density distribution from \citet{bruns05}
(darker regions represent higher $N($\hi$)$, see Br\"uns et al. for the detailed $N($\hi$)$ scale).
Our sightlines and major features of the Magellanic system are indicated. The box highlights 
a region known as the SMC Wing or Western end of the Magellanic Bridge.
\label{fig-map}}
\end{figure*}

In  Fig.~\ref{fig-map} we show the locations of our targets on an
\hi\ map of the Magellanic System from \citet{bruns05} (more properties 
about the stars can be found in Lehner 2002).
DI\,1388 and PKS\,0312--77 are situated approximately midway between the SMC and LMC, 
and DGIK 975 lies at the eastern end of the Bridge near the LMC halo, 
allowing us to probe different regions of the Bridge (see Fig.~\ref{fig-map}). 
All these sightlines are outside the SMC Wing that is approximately delimited
by the box in Fig.~\ref{fig-map}. 

The two stars were observed with {\em FUSE}, {\em HST}/STIS E140M, 
and the Anglo-Australian 3.9-m telescope (AAT). 
The QSO PKS\,0312--77 was observed with STIS 
E140M and {\em FUSE}, but the STIS and {\em FUSE}\ data are of too poor 
quality for a detailed metal-line absorption analysis. Unfortunately, the
{\em FUSE}\ program F018 (PI: Lehner) to obtain a good quality FUV spectrum 
of this QSO was not completed before the failure
of {\fuse}\ and only 16\% of the requested time was observed.  
However, the STIS data can be rebinned and used to study the Ly$\alpha$ absorption.  
For the STIS E140M and AAT observations of DGIK\,975 and DI\,1388, 
we refer the reader to \citet{lehner01} and \citet{lehner02}. The 
previous {\fuse}\ observations of these two stars are also described 
in \citet{lehner02}.  New {\em FUSE}\ observations 
were obtained for DI\,1388 (program U106 non-proprietary re-observations of science 
targets) and DGIK\,975 (program G050, PI: Lehner),
adding another 10 ks  and 29  ks exposure time, respectively. 
(Note that the total time for the program U106  is 
30 ks, but about half the exposures have no signal.) 
The total {\fuse}\ exposure times are therefore 28 ks for DI\,1388
and 52 ks for DGIK\,975.

All the {\fuse}\ data were recalibrated using the newest {\sc calfuse} version 
(v3.2, Dixon et al. 2007). The extracted spectra associated with the separate exposures 
were aligned by cross-correlating the positions of absorption lines, and then co-added.
The oversampled {\fuse}\ spectra were binned to a bin size of 0.027 \AA\ 
(4 pixels), providing about three samples per $\sim$20 \km\ resolution element. 
In order to achieve the optimum signal-to-noise, 
segments with overlapping wavelengths were coadded for the DGIK\,975
{\fuse}\ spectra. However, we ensured before coadding the various
spectra that none of the absorption lines of interest was affected
by fixed-pattern noise, by comparing the interstellar profiles in multiple
detector segments. 
The zero point in the final {\fuse}\ wavelength scale was
established by  shifting the average {\fuse}\ velocity to
the STIS 140M velocity of the same species (e.g., \oi, \nni, \feii). 
STIS data reductions provide an excellent wavelength calibration, with
a velocity uncertainty of $\sim$1 \km.

\section{Analysis}\label{sec-analysis}
\subsection{Metals}\label{sec-metal}
\begin{figure*}[tbp]
\epsscale{0.8} 
\plotone{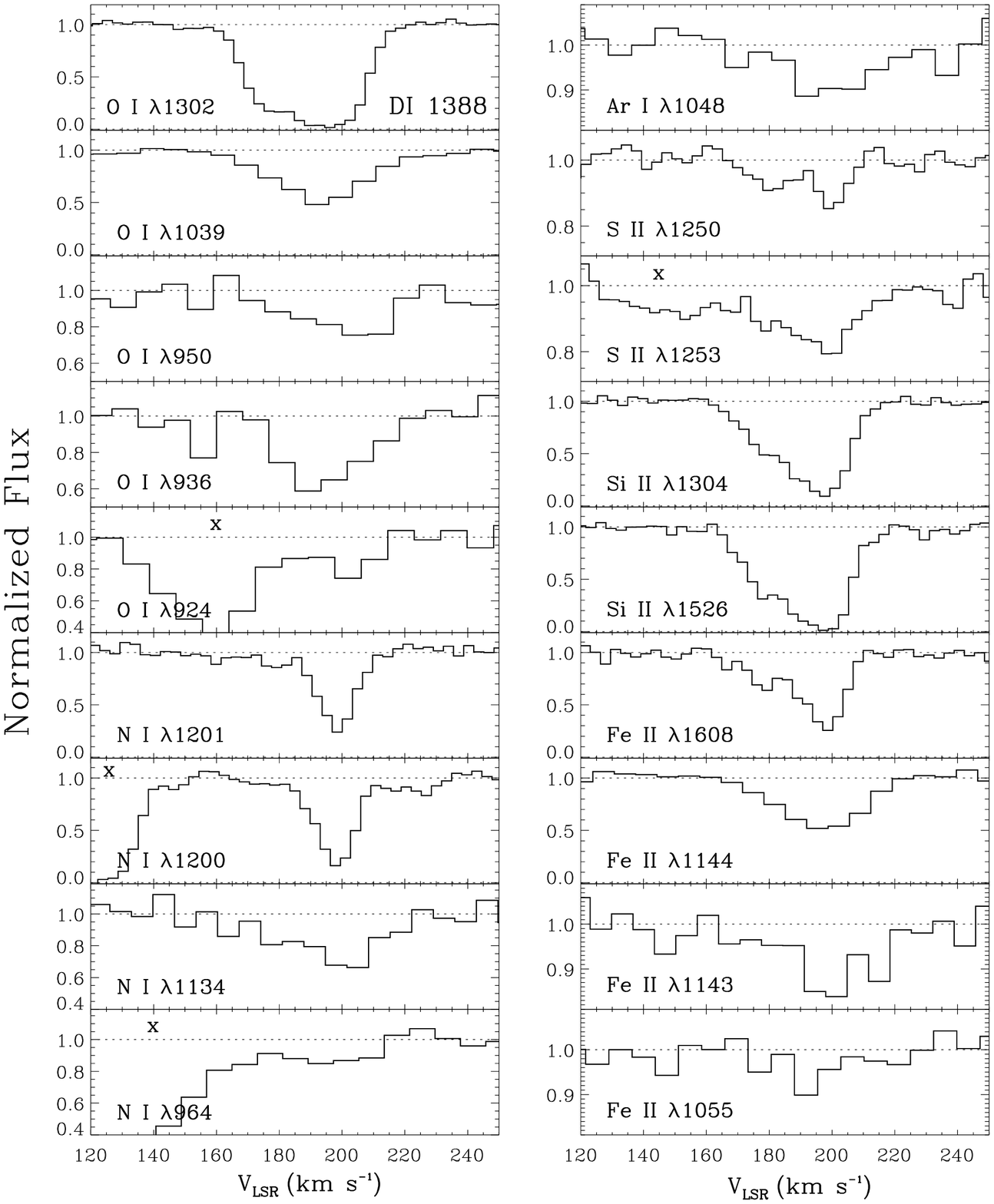}
\caption{Normalized profiles against the LSR velocity near the Bridge velocities of  
neutral and singly ionized species in the STIS E140M and {\fuse}\ spectra of DI\,1388.
The Bridge absorption occurs between about 160 and 220 \km\ toward this line of sight. 
The ``x'' shows part of the spectrum that is contaminated by other absorbing features. 
\label{fig-di}}
\end{figure*}

\begin{figure*}[tbp]
\epsscale{0.8} 
\plotone{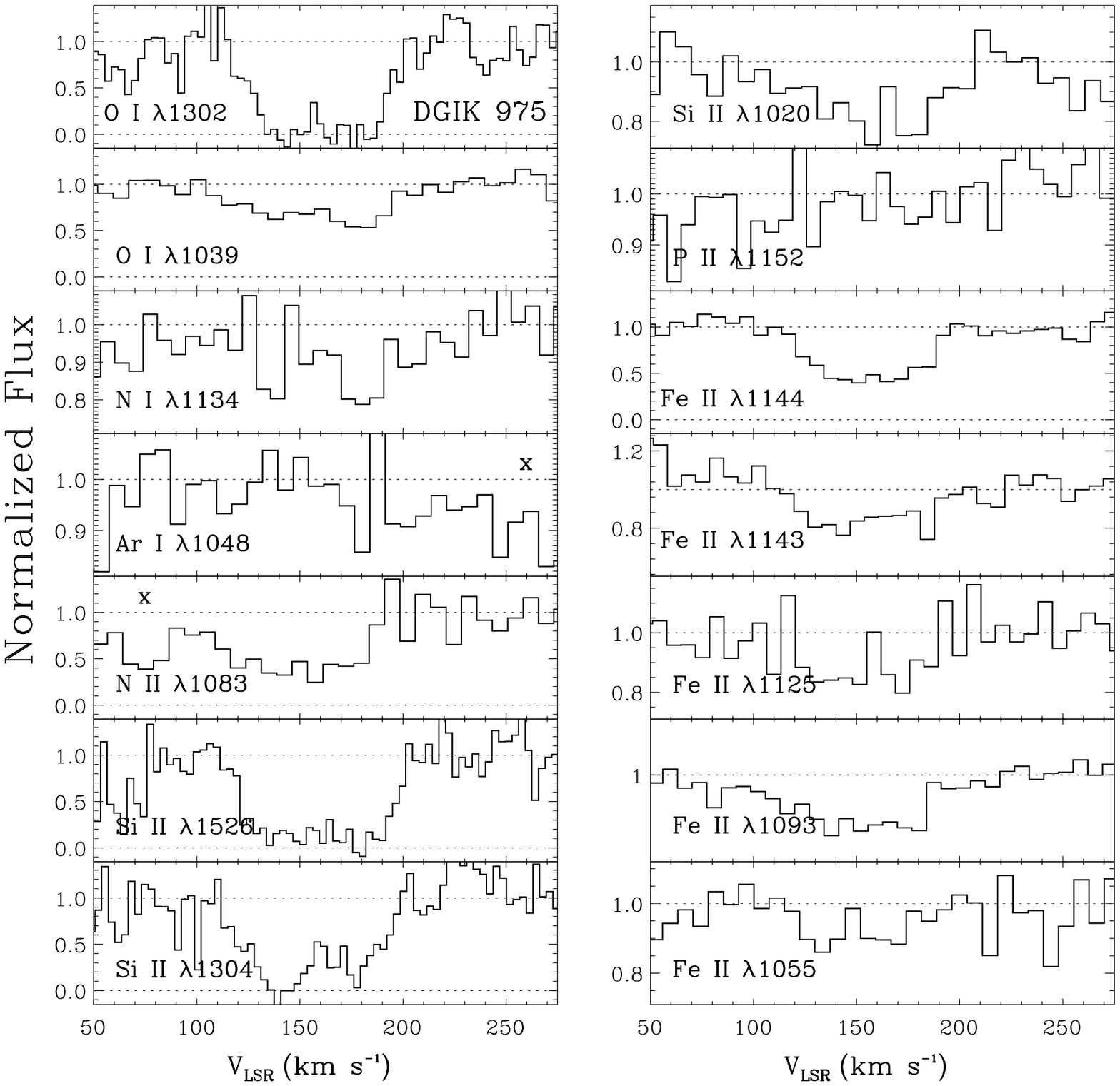}
\caption{Normalized profiles against the LSR velocity near the Bridge velocities of  
neutral and singly ionized species
in the STIS E140M and {\fuse}\ spectra of DGIK\,975.
The Bridge absorption occurs between about 100 and 200 \km\ toward this line of sight. 
The ``x'' shows part of the spectrum that is contaminated by other absorbing features. 
\label{fig-dgik}}
\end{figure*}

The continuum levels near the metal absorption lines 
were modeled by fitting Legendre polynomials within 
about $[-300,600]$ \km\ of each absorption line. Low-order polynomials 
were generally adopted, but in some cases high-order polynomials were
necessary (e.g. \feiii\ is blended with the stellar photospheric 
and wind lines).  For weak lines, several continuum placements were tested 
to be certain that the continuum error was robust \citep[see][]{sembach92}. 
In Figs.~\ref{fig-di} and \ref{fig-dgik}, we show the normalized spectra
for the interstellar metal-line transitions in DI\,1388 and DGIK\,975, respectively
\citep[see also][]{lehner01,lehner02}. Selected transitions have no serious blends with 
other features. 

In order to estimate the column density we adopted 
the apparent optical depth (AOD) method of \citet{savage91}. In this method 
the absorption profiles are converted into
apparent column densities per unit velocity $N_a(v) = 3.768\times
10^{14} \ln[F_c/F_{\rm obs}(v)]/(f\lambda)$ cm$^{-2}$\,(\km)$^{-1}$, where $F_c$ is the
continuum flux, $F_{\rm obs}(v)$ is the observed flux as a function of
velocity, $f$ is the oscillator strength of the absorption and
$\lambda$ is in \AA\ (atomic parameters were adopted from Morton 2003).  
The total column density was obtained by integrating over the absorption profile  $N_a= \int N_a(v)dv$.
According to \citet{savage91}, this method is adequate
for data with $b_{\rm line} \ga 0.25$--$0.50 b_{\rm inst}$, where $b_{\rm line}$
is the intrinsic $b$-value of the line and $b_{\rm instr}$ is the $b$-value 
of the instrument. Since $b \equiv$\,FWHM$/1.667$, for STIS E140M, $b_{\rm inst} \simeq 4$ \km\ 
and for {\em FUSE}, $b_{\rm inst} \approx 12$ \km. Therefore, we assume that
a negligible fraction of the gas has $b\ll 1$ \km, an assumption made implicitly in most 
ISM abundance analyses using these instruments. Yet, we note that if the apparent column densities of a species with similar 
transitions estimated by both intruments are similar, this implies that there must be
little or no unresolved saturation. 

The reader should be aware that signatures of cold gas in the Bridge have been found 
toward DI\,1388 and PKS\,0312--77. The fraction of cold gas with respect
to the warmer gas is unknown. We note that \citet{lehner02} derived $b \simeq 2.6$ \km\ 
for the H$_2$ in the Bridge toward DI\,1388 using a curve of growth with a 
single component. If a typical temperature of the molecular
gas is $T\sim 100$--200 K, turbulent motions would dominate the broadening of the 
H$_2$ lines. Hence even though there may be cold gas, turbulent motions may be
important enough to keep the broadening of the atomic and ionic lines greater than 1 \km. 
Because the investigated gas is multiphase \citep{lehner01,lehner02}, the AOD method 
is favored over the curve-of-growth (COG)  or profile fitting methods because the AOD does 
not make any a priori on the kinematical distribution of the gas along the sightline.
Finally, we note that our estimates are also based on non-detection of a line: these
strict limits are consistent with our AOD estimates (see below), given us confidence in 
the results presented here. 

When $\tau_a \ll 1$ (which is the case for several transitions used in
this work), unresolved saturation should not be problematic as long as $b$ is 
not much smaller than 1 \km. For stronger lines, unresolved saturated structure can 
be identified by comparing the lines of the same species with different $f\lambda$ . 
Following \citet{savage91}, the difference in $f\lambda$ must be a factor of 2 (or 0.3 dex) or more to be 
able to detect unresolved saturation. The $f\lambda$ values are summarized 
in Table~\ref{t-metal}. If some moderate saturation exists, we can
correct for it using the procedure described in \citet{savage91}. 
For various cases of blending and line broadening, 
they found a tight relation between the difference of the true column density
and the apparent column density of the weaker line 
versus the difference of the strong line and weak line apparent
column densities. The correction to the apparent column density
of the weak line for a given difference between  the strong- and weak-line 
apparent column densities are summarized in their Table~4. 

In cases where we have only reliable information on a single line, 
we only quote a lower limit on the apparent column density. However,
we note that due to the low metallicity of the Bridge, the peak apparent optical depths 
of metal lines are generally less than 1. Even strong transitions such as 
\oi\ $\lambda$1302 and \nii\ $\lambda$1083, which have line profiles that are 
usually completely saturated in the diffuse gas of Galactic or SMC/LMC environments 
with similar \hi\ column densities, do not reach zero flux (at least toward DI\,1388). 

When no  absorption is observed for a given species, we measured
the equivalent width (and $1\sigma$ error) over the same velocity range found
from similar species that are detected. The 3$\sigma$ upper limit on
the equivalent width is defined as the 1$\sigma$ error times three.
The 3$\sigma$ upper limit on the column density was then derived assuming
the absorption line lies on the linear part of the curve of the growth.

\begin{deluxetable}{lccc}
\tabcolsep=6pt
\tablecolumns{4}
\tablewidth{0pt} 
\tablecaption{Apparent Column Densities of the Metals \label{t-metal}} 
\tablehead{\colhead{Species}   &  \colhead{$\lambda$} &  \colhead{$\log f\lambda$} &\colhead{$\log N_a$}\\
\colhead{}   &  \colhead{(\AA)} &  \colhead{}&\colhead{}}
\startdata
  \cutinhead{DI\,1388}
\oi\   &  1302.168& 1.795 & $(>) 14.77 \pm 0.03$ \\
\oi\   &  1039.230& 0.974 & $(>) 14.97 \pm 0.06$ \\
\oi\   &  950.885 & 0.176 & $15.33 \pm 0.10$ \\
\oi\   &  936.630 & 0.534 & $15.26 \,^{+0.13}_{-0.16}$ \\
\oi\   &  924.950 & 0.154 & $15.32 \pm 0.13$ \\
\nni\  &  1200.710& 1.715 & $(>) 14.07 \pm 0.04$ \\
\nni\  &  1200.223& 2.018 & $(>) 13.89 \pm 0.04$ \\
\nni\  &  1134.980& 1.674 & $(>) 14.00 \pm 0.08$ \\
\nni\  &  964.626 & 0.882 & $(\le) 14.32 \,^{+0.13}_{-0.20}$ \\
\nni\  &  954.104 & 0.582 & $< 14.45$ \\
\ari\  &  1066.660& 1.857 & $< 13.17	   $ \\
\ari\  &  1048.220& 2.440 & $12.80 \pm 0.15$$^a$ \\
\siii\ &  1526.707& 2.308 & $14.14 \pm 0.07$ \\
\siii\ &  1304.370& 2.051 & $14.18 \pm 0.02$ \\
\sii\  &  1253.805& 1.136 & $14.25 \pm 0.10$ \\
\sii\  &  1250.578& 0.832 & $14.27 \pm 0.07$ \\
\feii\ &  1608.451& 1.970 & $13.95 \pm 0.02$ \\
\feii\ &  1144.939& 1.980 & $13.93 \pm 0.05$ \\
\feii\ &  1143.226& 1.341 & $13.95 \pm 0.10$ \\
\feii\ &  1055.262& 0.812 & $(\le)14.05 \,^{+0.15}_{-0.25}$ \\
  \cutinhead{DGIK\,975}
\oi\   &  1039.230& 0.974 & $(>) 15.20 \pm 0.04$ \\
\nni\  &  1134.980& 1.674 & $(\le) 13.88 \,^{+0.12}_{-0.16}$ \\
\ari\  &  1048.220& 2.440 & $<12.92$ 			\\
\siii\ &  1020.699& 1.225 & $(\le) 14.54\,^{+0.14}_{-0.20}$ \\
\feii\ &  1144.939& 1.980 & $14.18 \pm 0.03$ \\
\feii\ &  1143.226& 1.341 & $14.36 \pm 0.07$ \\
\feii\ &  1125.448& 1.245 & $14.44 \pm 0.10$ \\
\feii\ &  1093.877& 1.555 & $14.34 \pm 0.05$ \\
\feii\ &  1055.262& 0.812 & $(\le) 14.60 \,^{+0.10}_{-0.16}$ \\
\enddata
\tablecomments{Atomic parameters are from \citet{morton03}. $N_a$ is the apparent
column density in cm$^{-2}$, except when the ``$(<)$" sign is present, which is
a 3$\sigma$ upper limit; ``$(\le)$ indicates that the detection
is barely 3$\sigma$, and ``$(>)$" indicates that the line is likely saturated (see text 
for more details). 
$a$: We note that \citet{lehner02} found a column density twice smaller for \ari. 
N. Lehner revisited the original reduced data ({\sc calfuse} v2.0.5) and concludes
that the measurement was made using solely LiF\,1A, which appears to suffer from fixed-pattern
noise contamination (see Fig.~2 in Lehner 2002) but has been corrected with the new calibration. Indeed
the LiF\,1A and LiF\,2B segments reduced with {\sc calfuse} v3.2 give consistent results, 
and those are consistent with $N($\ari$)$ determined with LiF\,2B  {\sc calfuse} v2.0.5 data. 
}		
\end{deluxetable}

\begin{figure}[tbp]
\epsscale{0.8} 
\plotone{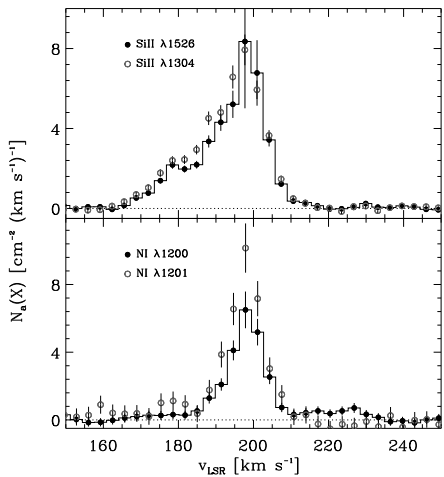}
\caption{Comparison of the apparent column density profiles for \siii\ 
and \nni\ for the DI\,1388 sightline. The \siii\ lines are essentially fully resolved while \nni\ shows 
some unresolved saturation. 
\label{fig-aod}}
\end{figure}

In Table~\ref{t-metal}, we present our estimates of the apparent column densities 
for DI\,1388 and DGIK\,975. Below, we review each sightline separately: 

{\em DI\,1388}: In Table~\ref{t-metal}, we present our raw
measurements of the apparent column densities for DI\,1388. 
Each considered species have at least 2 transitions with 
$\Delta \log(f\lambda) \ga 0.3$. 
\\
{\em Singly-ionized species:} In Fig.~\ref{fig-aod}, 
we show the apparent column density profiles of 
\siii\ $\lambda$$\lambda$1304, 1526 and the integrated
column densities agree remarkably well, especially 
at $\sim$198 \km\ where the absorption is the strongest. 
The continuum near \siii\ $\lambda$1526
is somewhat more complicated 
than for \siii\ $\lambda$1304, which results in a larger
uncertainty. A similar agreemeent is found for \sii, although
we note that  the continuum near the blue side of  
\sii\ $\lambda$1253 is more complicated. For \feii, the data are from {\fuse}\ 
and STIS. There is again an excellent agreement 
between the weak and strong transisitions. Furthermore, 
\feii\ $\lambda$$\lambda$1608 (STIS), 1144 ({\fuse}) have similar strengths
and apparent column densities. Therefore
the coarser spectral resolution of {\fuse}\ does not
seem to affect our apparent column estimates. 
Hence for the singly-ionized species, there is 
no evidence of saturation for the lines summarized in
Table~\ref{t-metal}. 

{\em \ari:} The transition at 1066.66 \AA\ is a 
3$\sigma$ upper limit. This is a firm upper limit
irrespective of the intrinsic broadening of the line. 
\ari\ $\lambda$1048 is detected and its 
apparent column density is consistent with 
the weaker transition.  If the strong transition
has some unresolved saturation, it is likely less that 0.1 dex.

{\em \nni:} The transitions at 1200.71 and 1134.98 \AA\
have similar strengths, but the agreement in the $N_a$ of 
these two lines is not as good as for \feii, even though 
the $N_a$ estimates of these two lines overlap within 
1$\sigma$. This suggests that there may be some unresolved
saturation. The apparent column density of \nni\ $\lambda$1200.22
is 0.18 dex smaller than \nni\ $\lambda$1200.711 (see also Fig.~\ref{fig-aod}). 
Therefore  $N_a($\nni\ $\lambda$1200.711$)$ needs
to be corrected for saturation. Using the results
of \citet{savage91}, we find that the correction
is 0.27 dex. The corrected column density of 
\nni\ is therefore $\log N($\nni$) = 14.34 \pm 0.08$ (where the 
errors include statistical and saturation
correction uncertainties).  This result can be directly tested with the weak
\nni\ $\lambda$964 transition, a 2.8--3.1$\sigma$ dectection, which
yields $\log N_a = 14.32 \,^{+0.13}_{-0.20}$ dex, in excellent agreement
with our above corrected column density (note that this \nni\ line has a similar
strength to \oi\ $\lambda$924 described below but
the continuum placement is much more straightforward for the \oi\ line, 
explaining the difference in the errors). 
The upper limit on \nni\ $\lambda$954 is also consistent with our estimate. 

{\em \oi:} The strong transitions at 1302.17 and 1039.23 \AA\ 
give much smaller apparent column densities than the weak
lines, and are therefore saturated. There are also 
two weak transitions at 924.95 and 950.89 \AA\ that
are 3.0--3.2$\sigma$ and 4.2--4.5$\sigma$ detections,
respectively.
Near the weakest transitions \oi\ $\lambda$924 
there is a contaminating feature on the blue side of 
the line, while there is a very weak contaminant on the
red side of \oi\ $\lambda$950 (see Fig.~\ref{fig-di}). 
However, the apparent column densities of \oi\ $\lambda$924 and 
$\lambda$950 are consistent with each other. We also tested this by
measuring half the \oi\ $\lambda$924 profile known to be 
free of contamination and 
the resulting $N_a \times 2$ is consistent with the value 
reported in Table~\ref{t-metal}.  The continua near these lines are  
well modeled with a straight line. 
\oi\ $\lambda$$\lambda$924,950 have  $\tau_a \ll 1$ 
and similar $ f \lambda$. \oi\ $\lambda$936 is about 0.3 dex stronger
than these lines and can be used to test for any unresolved structures. 
However, this line is partially blended with \hi\ $\lambda$937, 
complicating the continuum placement, which results in a larger
error. Within the errors, the apparent column densities of the 
three lines are consistent. Nevertheless, to be cautious,
we adopted the weighted average of the apparent column densities of 
\oi\ $\lambda$$\lambda$924,950 
and added an error in quadrature of $+$0.07 dex  to include any 
possible weak saturation.  Our adopted \oi\ column density is 
$\log N($\oi$) = 15.33 \,^{+0.11}_{-0.08}$. 

{\em DGIK\,975:} The STIS data are of lower quality than those for the DI\,1388
and the peak apparent optical depths are greater than 1 for all the detected species (see Fig.~\ref{fig-dgik}). 
We therefore use only the {\fuse} data. Unfortunately, the {\fuse}\ stellar spectrum
is not as well behaved as the one of DI\,1388, and we often have to rely 
on a single line. Stellar contamination and amount of \hi\  are more important 
(see \S\ref{sec-hi}), so that none of the weak \oi\ and \nni\ lines can be used. 
The $\lambda$1039 transition is the only available \oi\ line and is likely saturated,
providing a firm lower limit. (The apparent column density of \oi\ $\lambda$ cannot 
be estimated because the signal-to-noise is too low and its absorption reaches
zero-flux). \ari\ $\lambda$1048 is not detected and provides a firm 3$\sigma$ upper limit. 
\nni\ $\lambda$1134 is a $2.8$--$3.2 \sigma$ detection, and therefore the estimate of
the apparent column  should be considered as an upper limit. The same
applies for \siii\ $\lambda$1020 (the measurement of this line being complicated
by the uncertain continuum placement). Only for \feii, several transitions
can be measured. Following the above method, we estimate the 
average column density using the weak lines (i.e. excluding \feii\ $\lambda$1144), 
and add a systematic error to take into account possible unresolved structures. 

\begin{deluxetable}{lc}
\tabcolsep=6pt
\tablecolumns{2}
\tablewidth{0pt} 
\tablecaption{Adopted column densities of the Metals \label{t-adopt}} 
\tablehead{\colhead{Species}    &\colhead{$\log N$}\\
\colhead{}   &\colhead{}}
\startdata
  \cutinhead{DI\,1388}
\oi\   &   $15.33 \,^{+0.11}_{-0.08}$ \\
\nni\  &   $14.34 \pm 0.08$ \\
\ari\  &   $12.80 \pm 0.15$ \\
\siii\ &   $14.17 \pm 0.02$ \\
\sii\  &   $14.26 \pm 0.06$ \\
\feii\ &   $13.95 \pm 0.02$ \\
  \cutinhead{DGIK\,975}
\oi\   &   $>15.20$ \\
\nni\  &   $\le 13.88 \,^{+0.12}_{-0.16}$ \\
\ari\  &   $<12.92$ \\
\siii\ &   $\le 14.54\,^{+0.14}_{-0.20}$ \\
\feii\ &   $14.38 \,^{+0.12}_{-0.05}$ \\
\enddata
\tablecomments{See \S\ref{sec-metal} for more details.}		
\end{deluxetable}

Our adopted column densities for neutral and singly-ionized species 
are summarized in Table~\ref{t-adopt}. There is an overall agreement
with previous column density estimates \citep{lehner01,lehner02}, 
in particular if in \citet{lehner02} the \oi\ and \nni\ column densities estimated in the DI\,1388 spectrum 
are systematically smaller than presently, they nonetheless overlap within $1.5\sigma$. 
The main difference is that he relies on a single-component curve-of-growth
analysis using both weak and strong lines, which depends on an assumed velocity distribution 
that is likely more complex than one Gaussian component. We believe 
using the AOD method is a better approach to determine the most 
reliable column densities and errors in view of the complexity of the  velocity 
distribution in these sightlines.

\begin{figure*}[tbp]
\epsscale{0.9} 
\plotone{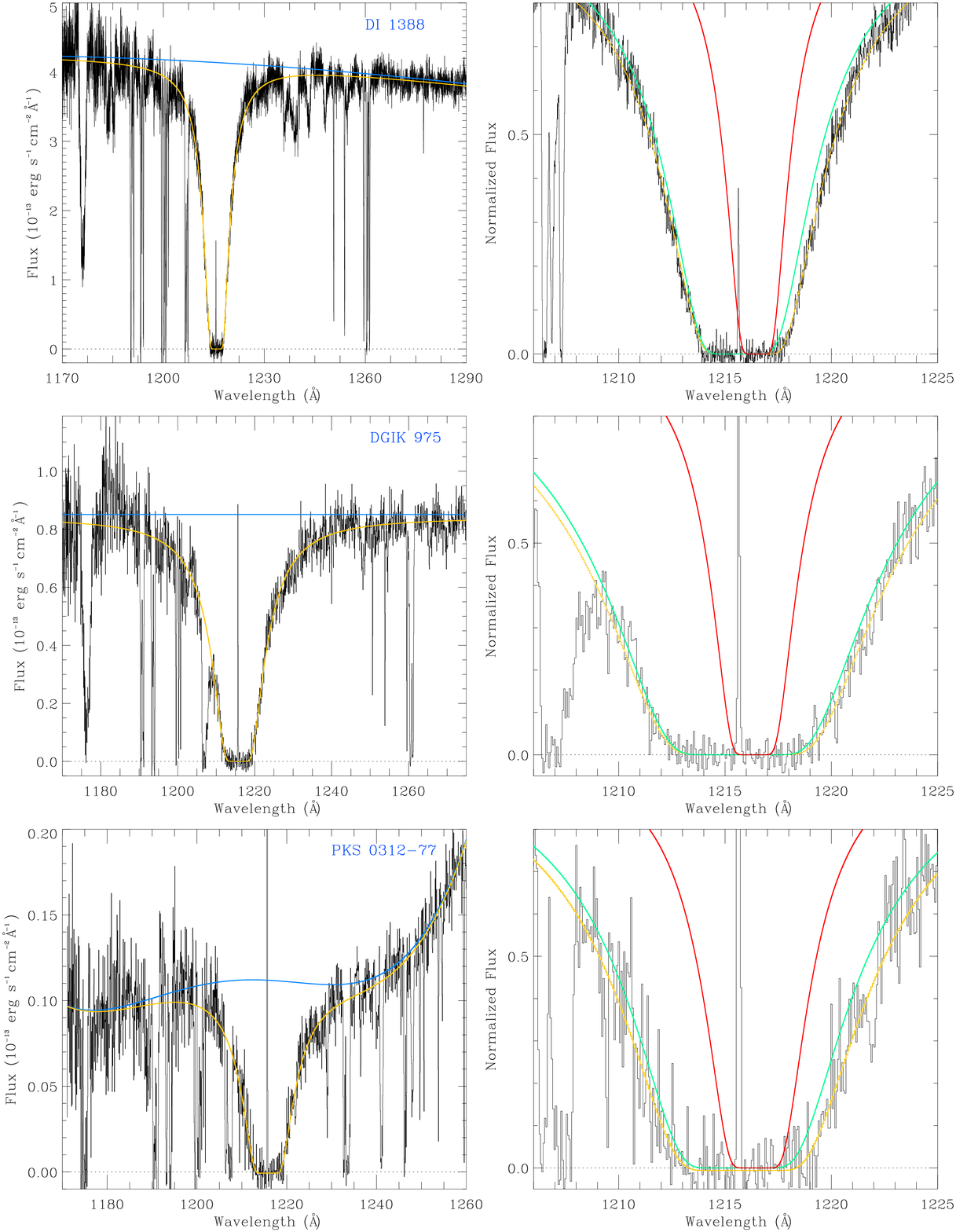}
\caption{{\em Left panels}: The \lya\ profiles in the STIS E140M spectra
of DI\,1388 ({\em top}), DGIK\,975 ({\em middle}), and PKS\,0312--77 ({\em bottom}). 
The blue line shows the continuum  and the yellow is the 
two-component fit to the interstellar \lya\ profile. {\em Right panels}: A zoom in on the core of the 
normalized \lya\ profiles is shown for each line of sight. The yellow line is the two-component
fit, while the green and red lines show only the Galactic and Bridge components, respectively. 
\label{fig-hi}}
\end{figure*}

\subsection{Neutral Hydrogen}\label{sec-hi}

\begin{deluxetable}{lcccc}
\tabcolsep=3pt
\tablecolumns{5}
\tablewidth{0pt} 
\tablecaption{\hi\ Column Density \label{t-hi}} 
\tablehead{\colhead{Sightline/}   &  \colhead{$v_{\rm LSR}$} &  \colhead{$\log N$}&\colhead{$\log N^\star$} &\colhead{$\log N_{\rm f}$} \\
\colhead{Component}   &  \colhead{(\km)} &  \colhead{}&\colhead{} &\colhead{}}
\startdata
DI\,1388 G  	& $-5$  & $20.41 \pm 0.02$		& \nodata	&  $20.41 \pm 0.02$  \\
DI\,1388 B  	& $195$ & $19.67 \pm 0.07$		& $18.58$	&  $19.63 \pm 0.07$   \\
DGIK\,975 G 	& $-5$  & $20.95 \pm 0.08$		& \nodata	&  $20.95 \pm 0.08$  \\
DGIK\,975 B 	& $160$ & $19.95 \pm 0.30$		& $18.98$	&  $19.90 \pm 0.30$   \\
PKS\,0312--77 G& $5$   & $20.78 \pm 0.06$		& \nodata	&  $20.78 \pm 0.06$  \\
PKS\,0312--77 B& $210$ & $20.12 \pm 0.30$		& \nodata	&  $20.12 \pm 0.30$   
\enddata
\tablecomments{``G": Galactic component; ``B": Bridge component 
The LSR velocities are fixed and are estimated from the \oi\ absorption
profiles. $N$ is the observed column density, $N^\star$ is the estimated
stellar \hi\ contribution for the H$\beta$ profiles (see text for more
details), and $N_{\rm f}$ was
corrected for the stellar component in the Bridge component. 
Note that the errors include different choices of continuum placement
and polynomial degree for the continuum, and velocity shifts of $\pm 5$ \km.}		
\end{deluxetable}

Here we discuss our derivation of the Bridge \hi\ column based
on the analysis of the \lya\ absorption.  The \hi\ column density 
is more complicated to derive because both the Galaxy and Bridge 
contribute to the absorption. \hi\ 21-cm emission observations
toward the stars are of little use for the Bridge component because 
the stars are embedded in the Bridge with an unknown depth. We therefore 
need to derive $N($\hi$)$ from the \lya\ absorption. To do so, we
fit each damped \lya\ absorption line profile with two components 
that correspond to the Galactic and Magellanic Bridge components.  
The blue-ward wing of the profile is due primarily to the 
Galactic component, but the Galactic component cannot completely 
account for the absorption in the red-ward wing, indicating the need 
for a Bridge component.  By fixing the velocities of the Galactic and 
Bridge components, we can determine the \hi\ column densities.
We note that the amount of \hi\ gas between the Galaxy and the Bridge is 
negligible, since the \hi\ 21-cm data do not show any emission toward these 
two lines of sight at a level of $10^{18.3}$ cm$^{-2}$, too small to affect 
the damping wings of \lya. \oi\ is the best metal proxy for \hi\ since its 
ionization potential and charge exchange reactions with hydrogen ensure 
that the ionization of \ion{H}{1} and \ion{O}{1} are strongly coupled. 
We can therefore use the kinematics of \oi\ to infer those for \hi. While 
the \oi\ profiles show multiple structures in both the Galactic and Bridge components, 
we only use the average velocities to fix the values for each component. 
These are listed in Table~\ref{t-hi}. Below we detail our fitting results for each star. 

{\it DI\,1388:} This line of sight is our best case because the continuum
of the star is well behaved and the Galactic \hi\ column density is not
so large that it fills the absorption in the red part of the \lya\ profile to a large degree. 
In  Fig.~\ref{fig-hi}, we show the profile of \lya\ with its continuum and  the 
fit with the two components (top-left panel). On the top-right panel, we show 
part of the  normalized \lya\ profile with the two-component fit in yellow, 
in red  the Bridge component, and in green the Milky Way component. This shows 
that while the Galactic component nearly fits the blue part of
the \lya\ absorption, there is extra absorption in the red part 
that is missed by the Galactic component. 
In Table~\ref{t-hi}, we summarize the derived
column density for each component in the third column.  
To test the robustness of our fit, we undertook many simulations where the velocity centroids were
changed by $\pm 5$ \km, the placement of the continuum was varied, 
and the continuum was modeled by a range of polynomials of degree 
of 1 to 4. The errors reported in Table~\ref{t-hi} were estimated
to reflect the 1$\sigma$ distribution between all the trials.  

{\it DGIK\,975}: This line of sight is more complicated because 
the Galactic \hi\ component is much stronger than
toward DI\,1388. For the latter, the Galactic \hi\ component is about
a factor 5 stronger than the Bridge component, while toward 
DGIK\,975, we found the Galactic component is $\sim$10 times stronger
than that of the Bridge.  The \siiii\ $\lambda$1206 line is also very 
strong and removes some information in the blue part of the
\lya\ absorption. Finally the STIS E140M DGIK\,975 spectrum has a much lower
S/N; we therefore rebinned these data by 5 pixels before fitting. In Fig.~\ref{fig-hi}, 
we show our resulting best fit overplotted on the \lya\ 
profile (middle-left panel), where the continuum is well behaved and 
can be modeled by a straight line.  The normalized profile (middle-right panel) indeed
shows that the difference between the two-component fit and the single Galactic
component fit is small, which results in a large error in $N($\hi$)$
of the Bridge feature. However, as for DI\,1388, at $ 1219 \la \lambda \la 1222$ \AA,
where the optical depth is more important, the Galactic component does not 
properly fit the red part of the absorption without the Bridge component.
We also undertook several trials in order to derive the 1$\sigma$ error reported
in Table~\ref{t-hi}. 

The \hi\ 21-cm emission data can help to evaluate the 
solution from our fits, at least for the Galactic component; indeed, since 
the whole column of Galactic gas is probed in both emission and absorption 
toward the stars, the infered column densities derived from absorption
and emission spectra can be compared. The main uncertainty that 
arises from such a comparison is that the beam that collected the \hi\ 21 cm 
emission data is much larger than the pencil-like beam of the absorption 
observations. Therefore the beam of the radio telescope only represents
an average column density of the region near the line-of-sight.  
\citet{wakker01} studied this effect for low-, intermediate-, and high-velocity 
clouds by comparing $N($\hi$)$ measured via a beam of 36\arcmin, 9\arcmin, 1\arcmin,
and through \lya\ absorption. They found that the ``beam-effect'' was 
less important for the low-velocity gas where the range of ratio of $N($\hi$)$
measured with a 9\arcmin\ versus a 36\arcmin\ is  0.9 to 1.4 than for the HVCs where 
the range of ratio is 0.3 to 2.1. This difference is because the physical sizes probed 
by the \hi\ 21 cm data are naturally smaller in nearby gas and therefore less likely to 
have large angular/spatial variations. Hence \hi\ 21-cm emission observations
observed through a 15\arcmin\ beam should provide reliable $N($\hi$)$ within a
beam error of about $\pm 0.10$ dex (see also Lehner et al. 2004)
for the low-velocity Galactic gas that can be compared to the results from 
the profile fitting of the \lya\ absorption.

With \hi\ 21-cm emission observations from the Parkes 64 m radio telescope with
a 15\arcmin\ beam  and 1 \km\ spectral resolution (M.E. Putman, 1999, private communication)
we derive the \hi\ column density for the Galactic component where we make the usual assumption 
that the medium is optically thin: $\log N($\hi$) = 20.44 \pm 0.11$ toward DI\,1388 and 
$\log N($\hi$) = 20.85 \pm 0.11$ toward DGIK\,975. Using  the 21-cm emission data 
from the 36\arcmin\ Leiden-Argentine-Bonn survey within $\sim$1\degr\ from the star's coordinates
\cite[LAB,][]{kalberla05,bajaja05}, we find 
Galactic \hi\ column densities that are systematically similar 
to those quoted above within the errors. Within $1\sigma$, 
the \hi\ column densities of the Galactic component derived from emission 
and absorption observations are consistent. 

Finally, we also consider  the QSO PKS\,0312--77 sightline:
since both the UV and radio observations probe the full depth of the Bridge, we 
can compare the resulting  \hi\ column densities measured with both observations with the 
caveat that we have to assume that small-scale variations in the 15\arcmin\ beam (corresponding 
to about 240 pc at 55 kpc) are small enough in the direction of  PKS\,0312--77.
\citet{kobulnicky99} detected \hi\ 21-cm emission for the Bridge component with the 
Parkes telescope toward this line-of-sight. They derived a Bridge \hi\ column 
density of $\log N($\hi$) = 20.09 \pm 0.13$, where the errors include 
statistical error and systematics associated with the beam that
is larger than our pencil beam for the UV observations (see above). 
We note that the LAB data give systematically larger $N($\hi$)$ for the Bridge 
but with relatively small variations within a radius (in the $l,b$ plane)
of $\sim$40\arcmin: the closest LAB pointing is 4\arcmin\ away where 
$\log N($\hi$) = 20.23 \pm 0.25$ and the range of $\log N($\hi$)$ is 
between 20.18 to 20.34 dex (toward the two stars, the variation in $N($\hi$)$
is quite larger -- a factor 2--3 -- within the same radius).   

The \lya\ profile of PKS\,0312--77 is shown in Fig.~\ref{fig-hi} (bottom-left panel) 
and is clearly more challenging to model than those for the two stars, 
as the S/N is lower (the data shown were rebinned by 5 pixels) and the red part
of the continuum is heavily affected by emission lines from 
the QSO (emission from \lyb, \ovi, and \oi), complicating 
the continuum placement. The Galactic
component is also strong, but because the whole Bridge gas is probed
along this line of sight, this results in a higher \hi\ Bridge column
density than toward the two stars. In Fig.~\ref{fig-hi}, we show one of our best fits, 
and as for the other lines of sight, the errors reflect many
trials. However, because of the complexity of the continuum, 
the degree of the polynomial was allowed to vary between 2
and 7; a 5$^{\rm th}$ degree polynomial is shown in the figure.  The Galactic component
again cannot account for absorption at $\lambda \ga 1219$ \AA.
Table~\ref{t-hi} summarizes the results and while the error is large, 
the measured value of $\log N($\hi$) = 20.12 \pm 0.30$ is quite consistent with the 
\hi\ 21-cm column density derived by \citet{kobulnicky99}. These authors do 
not provide $N($\hi$)$ for the Galactic component, but using the LAB data
we estimate that it is $\log N($\hi$) = 20.83 \pm 0.17$, consistent with the
\lya\ absorption estimate in the Galactic component. 

The general agreement between the $N($\hi$)$ estimates from
absorption and emission measurements gives further confidence
in our fits to the \lya\ absorption observed along each lines of sight. 
The error estimates were systematically estimated via several 
fit trials and we believe that we are conservative in our error 
estimates. The stellar sightlines need to be, however, corrected 
for  contributions from the stellar photospheres. 
We follow the method originally presented by \citet{diplas91} and expanded by 
\citet{howk99} to estimate the stellar \lya\ contribution, which involves measuring the 
photospheric H$\beta$ equivalent width of each star and using the
equation A7 in  Howk et al. to estimate $W($\lya$)$. The stellar \hi\ column density
density is then derived via $\log N($\hi$) = 18.27 + W($\lya$)$. 
Using our AAT spectra, we estimate that $W({\rm H\beta})= 2.1\pm 0.2$  \AA\ 
for  DGIK\,975 and  $W({\rm H\beta})= 2.1\pm 0.2$  \AA\ for DI\,1388. 
These correspond to stellar \hi\ logarithm column density of 
18.58 and 18.98 for DI\,1388 and DGIK\,975, respectively. The correction is larger
for DGIK\,975 because the temperature of the star is lower than 
the temperature of DI\,1388.  The last column 
of Table~\ref{t-hi} report the Bridge \hi\ column densities corrected
for the stellar contribution.

\section{Abundances and Physical Conditions in the Magellanic Bridge}\label{sec-result}
\subsection{Metallicity}
\begin{figure}[tbp]
\epsscale{1.0} 
\plotone{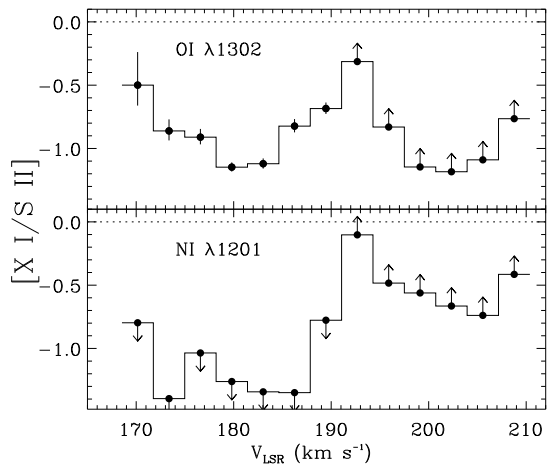}
\caption{Relative gas-phase abundance of \oi\ and \nni\ with respect to 
\sii\  (where $\lambda$1250) for the DI\,1388 sightline. At $v_{\rm LSR} \ga 193$ \km, 
\oi\ and \nni\ absorption suffers from saturation. 
\label{fig-aodcomp}}
\end{figure}

In Fig.~\ref{fig-aodcomp}, we show the ratio of the apparent column densities
of \oi\ and \nni\ to $N_a($\sii$)$ normalized to the solar abundances
for the DI\,1388 sightline. Sulphur is our reference because it is not 
depleted into dust. Also note that throughout the text we use 
the following notation ${\rm [X^i/Y^j]} = \log N({\rm X}^i)/N({\rm Y}^j) - \log({\rm X/Y})_\odot$, and we
adopt the solar abundances of \citet{asplund06}.\footnote{For the species used
throughout this work, the followings are the adopted solar abundances $\log ({\rm X/H})$: 
C: $-3.61 \pm 0.05$;
N: $-4.22 \pm 0.06$; O: $-3.34 \pm 0.05$; Si: $-4.49 \pm 0.02$; Ar: $-5.82 \pm 0.08$; 
S: $-4.85 \pm 0.03$; Fe: $-4.55 \pm 0.03$ \citep{asplund06}. Note that our error estimates
do not take into account the solar abundance errors.}
We explicitly retain the ionization state information to emphasize 
the impact of ionized gas on these measurements. As already discussed by \citet{lehner01}, 
the Bridge gas toward DI\,1388 has two main components, one mostly ionized 
($165 \le v_{\rm LSR} \le 193$ \km), the other one partially ionized ($193< v_{\rm LSR} \le 215$ \km).
The extremely ionized gas is illustrated by $[$\oi/\sii$] \simeq -1 $ for $170 \le v_{\rm LSR} \le 193$ \km\
(\sii\ is a important in both neutral and ionized gas, but \oi\ arises only in neutral gas).
Toward DGIK\,975, the normalized profiles shown in Fig.~\ref{fig-dgik} suggest a similar pattern: 
two main components are observed and the higher velocity component is more neutral than 
the lower velocity component as revealed by the stronger absorption in \oi\ and \nni\ and 
weaker \feii\ at 175 km and the opposite pattern is observed at $140$ \km. 
Therefore singly-ionized species cannot be directly used to estimate the metallicity
of the gas because they contain a significant contribution from ionized gas.

\begin{deluxetable}{lc}
\tabcolsep=6pt
\tablecolumns{2}
\tablewidth{0pt} 
\tablecaption{Gas-Phase Abundances of the Magellanic Bridge \label{t-abund}} 
\tablehead{\colhead{Species}  &\colhead{[X/H]} }
\startdata
  \cutinhead{DI\,1388}
\nni\ 	&  $-1.07 \pm 0.11 	  $	   \\
\oi\ 	&  $-0.96 \,^{+0.13}_{-0.11}	  $	   \\
\ari\ 	&  $-1.01 \,^{+0.16}_{-0.17}	 $	   \\
  \cutinhead{DGIK\,975}
\nni\ 	&  $(\le) -1.75\,^{+0.31}_{-0.38}$		    \\
\oi\ 	&  $>-1.36 (\pm 0.30)$		    \\
\ari\ 	&  $<-1.16 (\pm 0.30)$  	    
\enddata
\tablecomments{The adopted solar abundances are from Asplund et al. (2006): N: $-4.22$; 
O: $-3.34$; Ar: $-5.82$. The uncertainties in the solar abundances are not taken into
account in the errors listed in the table. A ``$<$" sign means a 3$\sigma$ upper limit. 
}		
\end{deluxetable}

On the other hand, the ionization of oxygen and hydrogen are strongly coupled, 
and  oxygen is only  mildly depleted into dust grains 
\citep[up to $-0.16$ dex using the updated \oi\ abundance -- see below --][]{meyer98}, implying 
that the \oi/\hi\ ratio is the best indicator of the metallicity of the Bridge gas.
We note that both the \hi\ column densities and molecular fraction 
in \citet{meyer98} are typically much larger than those 
in the Bridge gas. The oxygen measurements in the Local Bubble 
imply a depletion of $-0.12$ dex \citep{oliveira05}. 
Therefore the depletion of O in the Bridge is likely to 
be less than $-0.1$ dex.   We find
that the Bridge gas oxygen abundance is $[$\oi/\hi$] =-0.96 \,^{+0.13}_{-0.11}$ toward
DI\,1388. Toward  DGIK\,975,  only a lower limit is derived (see Table~\ref{t-abund}). 
We note that the solar abundance of O has changed quite significantly over the 
years and the new value may still be controversial. 
Prior measurements to \citet{asplund06} gave a solar O abundance $+0.17$ dex
larger \citep{grevesse98}, i.e  the Bridge O abundance would be further below. 

The solar Ar and N abundances appear more settled, with litte changes between 
those listed in \citet{grevesse98} and \citet{asplund06}. Furthermore,
Ar and N are not depleted into dust grains \citep{sofia98,meyer97}, and
\ari\ and \nni\  behave like \oi\ in gas with large \hi\ column density
\citep[see Fig.~6 in][]{lehner03}, i.e., the ionization fractions of these elements 
are coupled with the ionization fraction of H via a resonant charge-exchange
reaction. Toward DI\,1388, the values of the N and Ar abundances summarized in Table~\ref{t-abund}
show an excellent agreement with those of O.
This might appear puzzling at first glance since \oi\ is clearly detected in both components,
while \ari\ and \nni\ absorption is mostly absent in the 180 \km\ component. However, 
the neutral gas is clearly dominated by the component at 198 \km, since 
the column density in the 180 \km\ component is only about 10\% of the total
\oi\ column density (where we estimate $\log N($\oi$) \simeq 14.22$ in the 180 \km\
from a profile fit to the \oi\ $\lambda$$\lambda$1302, 1039, 963 lines).

The AOD estimate of \sii\ for $193 \le v_{\rm LSR} \le 215$ \km\ is 
$\log N($\sii$) = 13.98 \pm 0.08$. Assuming that 90\% of the
observed \hi\ is in this component (based on \oi) yields $[$\sii/\hi$] \simeq -0.8 $. 
Since absorption of \ciii, \nii, \siiii\ is found in this velocity range, 
this component is partially ionized. Therefore, the $[$\sii/\hi$]$ provides 
a strict upper limit to the abundance of S since the gas is partially 
ionized in this component. 

We mentioned above that O could be slightly depleted into
dust. If we assume an oxygen depletion of 0.05 dex (which is still consistent 
with the S abundance in Bridge and about 25\% ionized gas in this component), 
$[$\ari/\oi$] \sim -0.1$ and  $[$\nni/\oi$] \sim -0.2$. In low metallicity gas, 
N is generally found deficient \citep[e.g.,][]{henry07} 
(although we note for a metallicity of $-1$ dex, $[$N/O$]$ is more 
typically $-0.5$ dex), and this could explain the lower abundance of N. 
Nucleosynthesis history is unlikely to affect the Ar/O ratio since they
are both $\alpha$-elements. But \ari\ may be 
deficient because of photoionization \citep[see Fig.~6 in][]{lehner03}, although 
with $\log N($\hi$) = 19.6$ one would likely expect a deficiency less than 
0.1 dex (except possibly if a hard ionization source overionizes the edge of the 
neutral gas). (The \citet{grevesse98} value gives a large O depletion, but 
the $[$\nni/\oi$]$ would be less deficient). 
The errors can accommodate a slight O depletion (as compared with Ar and N) or none; 
we therefore use a weighted average of the abundances of \oi, \nni, \ari\ 
to derive the metallicity of the gas within the Bridge: $[{\rm Z/H}] = -1.02 \pm 0.07$.

Although the errors are large toward DGIK\,975, the limits on \ari\ and 
\nni\ suggest an extremely low metallicity as well. Using \ari\ and \oi\ (both 
$\alpha$-elements), we can bracket the metallicity of the Bridge gas toward DGIK\,975: 
$-1.7 < [{\rm Z/H}] < -0.9$. We note that toward this sightlines N may
appear more defficient than the $\alpha$-elements. 

Our gas-phase metallicity is in agreement with the metallicity
derived from 3 Bridge B-type stars, where the average is $-1.1 \pm 0.1$ dex  \citep{rolleston99,lee05}.
Since the  interstellar and stellar abundance measurements are made using different
techniques and have different systematics, the present-day metallicity of the Bridge is now 
secure with an average value  $[{\rm Z/H}]= -1.05 \pm 0.06$ dex. This is about a factor $\sim$3 times
smaller than the SMC abundance of $-0.6$ dex solar (see, e.g., Welty et al. 1997, Russel \& Dopita 1992, 
and references therein). In \S\ref{sec-discuss} we discuss the implication
of the metallicity results. 

\subsection{Ionization and Molecular Fraction}\label{sec-ionization}
Using the measurement of \hi\ along our two stellar sightlines, one can estimate the 
fractions of neutral, ionized, and molecular gas. Assuming an average 
Bridge metallicity from the stellar and interstellar measurements
of about $-1$ dex solar, we can use the singly and doubly ionized species
to estimate the amount of ionized gas. Toward DI\,1388, we use sulphur 
since this element is not depleted into dust and has 
the same nucleosynthetic history as
O and Ar since these are all $\alpha$-elements. 
Using the total column densities listed in Table~\ref{t-metal}, we 
find that the total hydrogen column density (H\,$=$\,\hi$+$\hii$+{\rm H_2}$) is 
$\log N({\rm H}) = \log[N($\si$)+N($\sii$)+N($\Siii$)] - \log (Z_{\rm MB}/Z_\odot) - \log({\rm S/H}_\odot)
\simeq 20.2$ dex (where we estimate $\log N($\Siii$)= 13.63\,^{+0.11}_{-0.15}$ from 
the line at 1012.495 \AA\ using the AOD method and $N($\si$)$ is negligible).
\citet{lehner02} derived $\log N({\rm H}_2) = 15.45$ toward DI\,1388. The fraction 
of neutral gas is therefore $f($\hi$) = N($\hi$)/N({\rm H}) = 0.27$, 
while the fraction of molecular gas is 
$f({\rm H}_2) = 2 N({\rm H}_2)/N({\rm H}) = 3.6 \times 10^{-5}$.
Therefore, about 70\% of the gas toward DI\,1388 is ionized. As we discussed 
above the lower velocity Bridge gas is nearly fully ionized ($\sim$95\%) while 
the higher velocity Bridge is partially ionized ($\sim$53\%, assuming O is not
depleted into dust and the O abundance from Asplund et al. 2006).

The near absence of \nni\ and \ari\ in the low velocity gas of the Bridge toward DI\,1388
can be mostly explained by photoionization. Indeed if steady 
photionization dominates, \ari\  is deficient because the
photoionization cross section of Ar is about 10 times that of H over
a broad range of energy. In our Galaxy, \ari\   has  been found to 
be deficient with respect to \oi\  \citep{sofia98,jenkins00,lehner03}.
In partially ionized gas, N also behaves more like Ar than O \citep{jenkins00,lehner03}.
Ionization is unlikely to affect the higher velocity gas because the 
$N($\hi$)$ is large enough to shield the gas against ionizing photons.  
Since there are hot early-type stars in the Bridge, 
a potential source for ionization is the photoionization by these objects.
In \S\ref{sec-discuss-ion} we discuss further the possible sources of ionization.
We also note that nucleosynthesis may also lower the N abundance. Using 
the lower limit on \nii\ ($>14.19$ dex) and the H column density derived
above, $[{\rm N/H}] > -1.51$ (\niii\ is negligible, see Lehner 2002). It is 
therefore not clear if N is deficient or not, although it appears less deficient
than usually observed for gas with a metallicity of 0.09 solar \citep[][and references therein]{henry07}.

Using the same arguments to that above for DGIK\,975 (and assuming a metallicity of 
$-1$ dex), but employing \feii\ 
and \feiii\ (where we estimate $\log N($\feiii$)= 14.29\pm 0.06$ from 
the line at 1122.524 \AA\ and assume that \feiii\ is not contaminated by the star) 
and correcting for the depletion of Fe (where we assume
it is $-0.6$ dex based on the similarity of the $[$\feii/\siii$]$ toward
both stellar sightlines and that the same depletion applies for 
\feii\ and \feiii), we find 
$\log N($H$) \sim 20.8$, which implies that $\sim$75--95\% of the gas is also 
ionized along this line of sight. 
No H$_2$ has been found  toward this sightline, implying 
$f[{\rm H}_2]\la  5\times 10^{-6}$.
As for DI\,1388, the low velocity component apppears 
more ionized than the higher velocity cloud (see above), but the Bridge 
gas toward DGIK\,975 appears even more ionized than along the DI\,1388 sightline.

Toward PKS\,0312--77, we estimate via the AOD method
$\log N($\sii$) = 14.95 \pm 0.07$ using the \sii\ $\lambda$$\lambda$1250,1253 lines. 
From the \hi\ column density derived by \citet{kobulnicky99}, 
we find that  $[$\sii/\hi$] = -0.29 \pm 0.16$. 
Assuming that the metallicity is $-1$ dex along this sightline,
this implies again that about 70--85\% of the gas 
is ionized toward PKS\,0312--77. We note that this line 
of sight probes the whole depth of the Bridge and therefore 
shows that ionized gas appears significant everywhere. 
We also note that none of these estimates take into account the highly ionized
phase probed by \ovi, \siiv, and \civ\ absorption \citep[see][]{lehner01,lehner02}. 
The weakly and highly ionized gas  are unlikely to probe the same material, 
and therefore these ionized fractions should be considered as lower limits.

\subsection{\cii*\ Diagnostic}
\begin{figure}[tbp]
\epsscale{1.0} 
\plotone{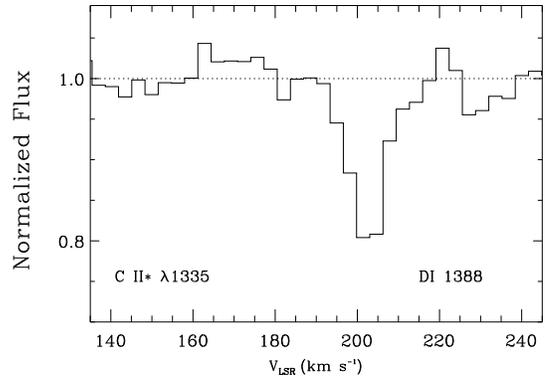}
\caption{Normalized profile of \cii*\ against the LSR velocity. Note
the absence of absorption between $\sim$160 and 190 \km\ where the nearly
fully ionized component is observed. 
\label{fig-cii}}
\end{figure}

\cii*\ can be used to estimate the electron density and radiative cooling rate 
of the gas \citep[see e.g.][]{lehner04}. In ionized gas and cold neutral gas, 
the \cii\ radiation is a far more important coolant than in warm neutral gas \citep{wolfire95}.
Toward DI\,1388, \cii*\ is detected between 193 and 215 \km, i.e 
in the partially ionized component, but is absent between 165 and 193 \km, 
i.e. in the nearly fully ionized component  (see Fig.~\ref{fig-cii}).
We  find $\log N($\cii*$) = 12.77 \pm 0.11$ with an average velocity of 
$v_{\rm LSR} = 202.8 \pm 1.5$ \km and a $b$-value $b = 4.2\pm 1.1$ \km\ (obtained 
from a single-component profile fit), which implies $T<2\times 10^4$ K. 
It is therefore not clear if \cii*\ arises in some ionized gas at $T\sim 10^4$ K or
in cold neutral gas where the turbulent motions may be important.  
For the gas observed at $165 \le v_{\rm LSR} \le 193$ \km,  we derive 
a 3$\sigma$ upper limit: $\log N($\cii*$) < 12.18$. 
Toward DGIK\,975, the 3$\sigma$ upper limit is 13.4 and is not useful.

If we assume that the electron collisions dominate the excitation of 
\cii, 
the electron density  within the Bridge can be written as  \citep[see][]{lehner04}:
\begin{equation}\label{elredeqt}
n_e  \simeq 0.183 \sqrt{T} \,\frac{N({\rm C}^{+*})}{N({\rm C}^{+})} \, {\rm cm}^{-3}\,.
\end{equation}
We assume a temperature of $T \sim 10^4$ K \citep[typically observed in ionized gas, e.g.,][]{howk06}.
We use \sii\ as a proxy of \cii\ and assume that carbon is 0.25 dex depleted 
\citep[][and using the updated Asplund et al. 2006's solar C abundance]{cardelli96}. 
For gas at  $165 \le v_{\rm LSR} \le 193$ \km, we find $\log N($\sii$) = 13.94 \pm 0.08$
and therefore  $n_e < 0.03$ cm$^{-3}$ (3$\sigma$) at $T = 10^4$ K. Since this gas is nearly
fully ionized, $n_e \approx n_p$, and therefore the overall density of the ionized
gas is quite low. From \sii, $\log N($H$) \sim 19.8$ (including \Siii\ would increase
this column by at most 0.2 dex); the pathlength of probed ionized gas is $>$0.6 kpc. 
Since the projected size of the Bridge is about 5 kpc, DI\,1388 is likely not 
deeply embedded in the Bridge (this is consistent with the comparison of the column densities 
from \hi\ emission and \hi\ absorption). For the gas at  $193 \le v_{\rm LSR} \le 215$ \km, we find  
$n_e < 0.1$ cm$^{-3}$ at $T = 10^4$ K (here it is an upper limit because of the 
\cii*\ absorption could possibly arise in cold gas). 

Since the H$\alpha$ intensity scales with the square of the density, it is not surprising 
that most of the current H$\alpha$ observations of the Bridge outside dense \hii\ regions  yield no  
H$\alpha$ detection down to $ I[{\rm H}\alpha]  < 0.5$--2 R \citep{putman03,muller07}. 
\citet{lehner04} estimated for the Galactic warm ionized medium (WIM) the relationhip between \cii*\ 
produced in the WIM and the intensity of H$\alpha$. Adapting their relationship (Eqt. 7 in their paper) 
to the Bridge conditions,  $I[{\rm H}\alpha] \simeq 1.5 \times 10^{-13}\, N_{\rm WIM}($\cii*$)$ Rayleigh
(where $N_{\rm WIM}($\cii*$)$ is in cm$^{-2}$). With our 3$\sigma$ upper limit $\log N($\cii*$) < 12.18$, 
we derive $I[{\rm H}\alpha] < 0.2$ R, typically smaller than current limits from H$\alpha$
observations. In the 
partially ionized component $ I[{\rm H}\alpha]  < 0.9$ R. Future deep H$\alpha$ observations
with the Wisconsin H-Alpha Mapper (WHAM) currently being installed in Australia should yield important 
clues on  the faint, diffuse H$\alpha$ emission in the Bridge and other tidal structures around 
the Magellanic Clouds.

The \cii\ radiative cooling rate is expressed as \citep[see][]{lehner04}:
\begin{equation}\label{ecool2}
l_c   = 2.89 \times 10^{-20}\frac{N({\rm C}^{+*})}{N({\rm H})}\, \,{\rm erg}\, {\rm s}^{-1}\,  ({\rm H})^{-1}\,.
\end{equation}
As discussed the \cii*\ absorption is only detected in the partially ionized gas. 
Most of the \hi\ is contained in this component and therefore the \cii\ radiative cooling rate
is $4.5 \times 10^{-27}$ erg\,s$^{-1}$\,(\hi)$^{-1}$. We can express this as the cooling 
per nucleon if we use the amount of 
\sii\ present in this component: we derive $l_c < 2.8 \times 10^{-27}$ erg\,s$^{-1}$\,(H)$^{-1}$
(it is an upper limit because we neglected \Siii). 
Since the  cooling rate is directly proportional to the metallicity of the gas for ionized
material \citep[e.g.,][]{wolfire03}, 
these cooling rates per metal are similar to the Galactic average rate within the 1$\sigma$ dispersion 
derived by \citet{lehner04} at similar $N($\hi$)$ or $N({\rm H})$. Where we do not 
detect \cii*, $l_c < 7 \times 10^{-28}$ erg\,s$^{-1}$\,(H)$^{-1}$, much smaller than 
those observed in the Milky Way at low $N($\hi$)$.

\section{Discussion}\label{sec-discuss}
In Table~\ref{t-prop}, we summarize the properties of the Bridge derived from this work. 
The most recent estimate of the \hi\ mass of the Bridge was derived by \citet{bruns05}, and to estimate the mass of 
H of the Bridge we assume that the large ionization fraction in the Bridge observed in 
our 3 lines of sight is characteristic of the Bridge as a whole.  Since our sightlines probe very 
different regions of the Bridge, the hypothesis that the ionization fraction 
is large in most regions of the Bridge appears reasonable. This mass is  $\sim$5 times
larger than usually estimated in current models \citep[e.g.][]{muller07a}.
However, since the gas mass fraction of the SMC and LMC and the ratio of the gas disk 
to the stellar disk of SMC can be adjusted in models, and since the SMC mass
is not known to better than a factor two, it is possible that simulations 
may be able to accommodate a larger Bridge mass.

\begin{deluxetable}{lc}
\tabcolsep=6pt
\tablecolumns{2}
\tablewidth{0pt} 
\tablecaption{Summary  of the Properties in the Diffuse Gas of the Magellanic Bridge \label{t-prop}} 
\tablehead{\colhead{Parameter}    &  \colhead{Value} }
\startdata
$[{\rm Z/H}]$$^a$		    & $-1.05 \pm 0.06 $  \\
$[$\ovi$/$H$]$			    & $-3.8$ to $-3.3$  \\
$[$\ovi$/$\hi$]$		    & $-3.1$ to $-2.3$  \\
Depletion of Si			    & $-0.45 \pm 0.06$  \\
Depletion of Fe			    & $-0.61 \pm 0.06$  \\
Fraction of total \hi\		    & $\sim$20\%   \\
Fraction of total \hii\		    & $\sim$80\%   \\
Fraction of H$_2$		    & $\la 0.004$\% \\
Density $n_e \simeq n_p$$^b$	    & $<0.03$--$0.1\sqrt{T_4}$  cm$^{-3}$ \\
Total H mass$^c$		    & $\sim 9\times 10^8$ M$_\odot$
\enddata
\tablecomments{$a$: Average metallicity from the interstellar and B-type stellar estimates.
$b$: Density of ionized gas. $T_4$ is the temperature in units of $10^4$ K. 
$c$: Adopting a total \hi\ mass of $1.8 \times 10^8$ M$_\odot$ \citep{bruns05} 
and assuming that the fraction of ionized is 80\% throughout the Bridge.   
Note that the {\em total} mass of the Bridge should include the H$+$He mass. 
}		
\end{deluxetable}

\subsection{Sources of Ionization of the Magellanic Bridge}\label{sec-discuss-ion}
One likely source of ionization of the Bridge is the stars within
the Bridge. A young population of stars was discovered 
throughout the Bridge \citep{irwin90,bica95,demers98}. The density of OB associations
is, however, not uniform: the OB-type stars are highly concentrated 
in the SMC Wing where the \hi\ column densities are the largest 
(roughly delimited by the box in Fig.~\ref{fig-map}) and are far 
more sparse farther away from the SMC \citep{irwin90,battinelli92}. The rectangle 
shown in Fig.~\ref{fig-map} also corresponds to the CO survey undertaken 
by \citet{mizuno06}, where CO emission was discovered, suggesting that
conditions might be conducive to star formation.  We note that the star-formation
history must be quite different from the less dense region of Bridge (see \S\ref{sec-per}), 
since in the SMC Wing, an SMC-like metallicity is derived from the
analysis of B-type stars \citep{lee05}.\footnote{There might be 
some confusion in the literature with the nomenclature
``SMC Wing'' and ``Bridge'' (also called InterCloud Region -- ICR). In some cases, the
Wing is an environment by itself (usually in the studies
of stars), in other cases (usually in  the studies of CO and \hi\ emission) 
it is a direct component of the Magellanic Bridge. The latter
is possible and seems to be justified in view of the \hi\ column density and velocity maps. 
But  the SMC Wing has a much higher concentration of OB associations \citep{irwin90}
and an SMC-like metallicity \citep{lee05}, which seem to imply a different evolution from the
Bridge at RA(J2000)\,$\ga 2^{\rm h}\,30^{\rm m}$ (see \S\ref{sec-per}). In particular, in view of these 
differences, it does not seem to be justified to extrapolate the properties derived
in the SMC Wing to the whole Bridge and vice versa, at least not before a better understanding
of the connection between these two regions.} 
Yet, the presence of early-type stars of age 20 Myr old or younger well inside and throughout the
Bridge \citep{demers98,rolleston99} implies ongoing star-formation at some level. 

The H$\alpha$ intensity of a cloud can be used to estimate 
the incident Lyman continuum flux. In our Galaxy, \citet{reynolds84,reynolds93} 
has used this approach to calculate the power required to ionize
the Galactic warm ionized medium up to 2--3 kpc above the plane, comparing it with the power available
from OB-type stars. Following \citet{tufte98}, the incident Lyman continuum flux is
$F_{\rm LC} = 2.1\times 10^5 I[{\rm H}\alpha]$ photons\,s$^{-1}$\,cm$^{-2}$, where 
$I[{\rm H}\alpha]$  in Rayleigh is assumed to arise solely from photoionization. 
Using the 3$\sigma$ upper limit derived from \cii*, we find for the diffuse ionized gas of 
the Bridge (outside denser \hii\ regions around OB-type stars) 
$F_{\rm LC}(\rm MB) < 0.4 \times 10^5$ photons\,s$^{-1}$\,cm$^{-2}$.
This rate is at least 100 times smaller than the one found in 
the Galaxy at high galactic latitudes \citep{reynolds84}. 
This rate would increase by a factor four if we used 
the \cii*\ detection but as discussed above the origin
of the excitation is uncertain as both cold neutral and warm 
ionized could contribute to the observed absorption. We therefore
concentrate in this section on the origin of the nearly fully ionized
gas. 

The flux of Lyman continuum photons from stars is highly dependent
on their types \citep{panagia73}. For example, an O6-type star will produce
a factor 10 more ionizing photons than an O9. But more importantly
the rate of photons drops dramatically for B1 and later types: a B1 compared
to an O9 produces 600 times fewer ionizing photons per second. In our Galaxy, 
surveys of OB-type stars have shown that O-type stars produce 
93\% of ionizing photons, even though they are much less numerous than
B-type stars \citep{terzian74,abbott82}. 

Unfortunately, little is known about the OB-type distribution in the Bridge, 
especially outside the SMC Wing. \citet{irwin90} reported an O8 star outside
the SMC Wing, and therefore there is likely to be O-type stars in the Bridge. 
In the SMC Wing, \citet{pierre86} estimated that the number of ionizing
photons over a very small area (0.1 square degree) is 
$ 70 \times 10^6$ photons\,s$^{-1}$\,cm$^{-2}$. Such a value would only require
0.1\% escaping Lyman continuum photons from the SMC Wing to ionize the Bridge. 
However, the sample of \citet{pierre86} is small and may be not be representative
of the OB distribution in the Wing. In particular, 20\% of the stars in their sample with 
types earlier than B0 are O7, which seems quite a larger number (in our Galaxy
this number is about 4\%, e.g., Terzian 1974); the three O7 stars in their sample 
contribute to the majority  of ionizing photons. A better characterization of 
the spectral types of the stars in the SMC Wing and Bridge would clearly help 
to better understand this important source of ionizing photons.

OB stars in the Bridge and Wing will also create an  environment that is favorable 
for the escape of Lyman continuum photons and for sustaining a high level 
of ionization throughout the Bridge. Indeed, OB associations through the winds
and death of massive stars can produce large bubbles and chimneys that can help ionizing
photons to travel large distances \citep[e.g.,][]{dove94,dove00,norman89}.
Such \hi\ shells and even possibly blow-outs and chimneys were surveyed by
\citet{muller03}, but only in the Wing at RA\,$< 3^{\rm h}$ and some 
may be only due to projection effect \citet{muller07a}. 
Furthermore, in an overall low density medium, it is likely that signatures of supernova
remnants and stellar winds will be difficult to decipher since rather than forming 
shells as observed in dense \hi\ regions, they might just dilute themselves in their surroundings.

Supernovae could also play a role in the ionization of the Bridge. 
Since stellar formation has occured for  200--300 Myr \citep{harris07}, 
it is likely that there have been several supernovae in and near the Bridge.
Assuming steady ionization of H (with $\sim$13.6 eV per ionization), our limits
on H$\alpha$ require $< 9 \times 10^{-7}$ erg\,s$^{-1}$\,cm$^{-2}$. Over
the entire Bridge (assuming a thickness of 5 kpc), this corresponds to a constant
power input to the gas of $<7\times 10^{38}$ erg\,s$^{-1}$. This power requires 
a supernova rate of about one every 50,000 yr  \citep[see][]{abbott82}. The supernova
rate is unknown and, as discussed by \citet{reynolds84}, only a small fraction ($\sim 17$\%) of the kinetic 
energy injected in the interstellar medium may be converted into ionizing hydrogen. 
Yet the presence of young, massive stars, supernovae have likely contributed to the ionization
and heating of the Bridge. 

Highly ionized gas found in the Bridge \citep{lehner01,lehner02} is  an indirect signature 
of hot gas that is still present in significant amounts (e.g. where the observed high ions are in an 
interface between the hot and cooler gas) or that was present and is in the process of
cooling. Both interfaces and cooling of hot gas can produce photons that may be important 
sources of photoionization of the environment \citep{slavin00,borkowski00,knauth03}. 
For example, \citet{slavin00} modeled the contribution to the photoionization from 
cooling of hot supernova remnants in the disk of our Galaxy and found that this source was adequate 
to account for the observed ionization of the Galactic halo. Therefore supernovae in the
Bridge and in the SMC and LMC near the Bridge could not only inject energy and heat the Bridge
but also their cooling remnants may provide a source of ionization. 

While the escaping of ionizing photons from the SMC Wing could
be an important source of ionization, 
the escape of ionizing photons from the Galactic, LMC, and SMC disks \citep[see][]{bm99,dove00} 
is also likely to contribute to the ionization of the Bridge.
Figure~8 in \citet{putman03} shows predicted $I[{\rm H}\alpha]$ near the LMC 
caused by the escape of ionizing photons from the LMC and the Galactic stellar bulge. 
The H$\alpha$ emission estimate produced from the leaking photons of the LMC (and Galaxy, 
but the latter is rather negligible) could reach 0.1 to  0.25 R near the LMC. 
Thus, such leakage alone is likely an  important source of ionizing photons for 
the Bridge.

While it is not clear which ionizing source is 
dominant in the Bridge, there appear to be enough photons to maintain
the high level of ionization seen in the Bridge. The presence of massive stars in the rarefied environment
of the Bridge is likely sufficient in itself for this environment to be 
largely ionized. Since cooling and recombination times are 
expected to be longer than in the Galactic environment because the Bridge gas has 
an extremely low metallicity and density, the ionization is likely to be sustainable for 
a long time. For example, for a $10^6$ K gas with a density $10^{-3}$ cm$^{-3}$, the radiative cooling is 
$t_{\rm cool} \sim 2$--$3\times 10^9$ yr for a 0.1 solar metallicity \citep{gnat07}, 
about 10 times longer than for a solar metallicity environment.  
Furthermore, according to \citet{harris07}, star formation has occured in the Bridge for 
the last 200--300 Myr, with a lower rate over the last $\sim$40 Myr. 
Therefore, several generations of OB stars could have ionized
the Bridge over a long time period.

We finally note that the ionization fraction toward DGIK\,975 is not only larger than 
toward DI\,1388, but the gas is also more highly-ionized. \citet{lehner02}
reported the measurements of \ovi, \siiv, and \civ\ toward these stars. In particular, 
$N_{\rm DGIK\,975}($\ovi$) \simeq 13 N_{\rm DI\,1388}($\ovi$)$, while 
$N_{\rm DGIK\,975}($\siii$) \simeq 2 N_{\rm DI\,1388}($\siii$)$. 
One can refer to Fig.~11 in \citet{lehner02} to see the dramatic
difference in the \ovi\ absorption profiles toward DI\,1388 and DGIK\,975. 
Recently, \citet{lehner07} argued for the presence of outflows
and even possibly a hot galactic wind from the LMC. 
Since DGIK\,975 is in the outskirts of the LMC and DI\,1388 is
farther away from the LMC (see Fig.~\ref{fig-map}), 
the larger amount of ambient hot gas toward DGIK\,975 might be related to 
the ouflows from the LMC. Perhaps the feedback processes occurring in the 
LMC inject energy into the Bridge. We note that the amount of metals incorporated into the 
Bridge gas from the LMC at large distance must be extremely small 
since neither the interstellar nor stellar abundances 
\citep[$-1.2 \pm 0.2$ dex solar for DGIK\,975, see][]{rolleston99,lee05} suggest an increase of 
the metallicity in this region.

\subsection{Metallicity of the Magellanic Bridge}

The metallicity of the Bridge is difficult to accommodate
with most models of the SMC-LMC-Galaxy interactions (but see below). 
The present-day low metallicity of the Bridge has, however, a simple explanation 
in the context of the chemical evolution of the SMC
if the Bridge is much older than 200 Myr. Indeed the bursting model for 
the star formation rate in the SMC described by \citet{pagel98} 
suggests that its metallicity only slowly increased from $-1.4$ to $-1.1$
dex between about 12 and 2.5 Gyr ago. In this model \citep[see also][]{harris04,idiart07}, 
it is only in the last $\sim$2.5 Gyr that the 
SMC metallicity has exceeded the modern Bridge value.
This is a large window for forming the Bridge from SMC gas, and suggests that the
Magellanic Bridge could be 10 times older than current  N-body simulations suggest.
\citet{harris07} showed that there were several bursts (2.5, 0.4, 0.06 Gyr ago) 
of star formation in the SMC in the last  $\sim$3 Gyr, but the most important one being 2.5 Gyr ago. 
New bursts of star-formation are believed to be often caused by the effects of close 
tidal interactions between galaxies \citep[e.g.,][]{larson78,barton07}. 
While \citet{harris07} noted that the bursts at 2.5 Gyr and 0.4 Gyr are temporally
coincident with past perigalactic passages of the SMC and Galaxy, these epochs also 
correspond to close encounters between the SMC and LMC \citep{kallivayalil06}. 
Due to its proximity, the LMC tidal perturbations on the SMC are larger
than those of the Galaxy \citep{lin95}. However, calculations of the orbits
of the SMC and LMC with the updated proper motions of the SMC and 
LMC still predict two more perigalactic approaches between the SMC and LMC since the 2.2 Gyr close encounter 
at $\sim$1.4, 0.2 Gyr, the latter being the closest approach \citep{kallivayalil06}. 
It is therefore unclear if the Bridge could be stable over 2--3 Gyr
(Gardiner \& Noguchi 1994 argued it could not be stable over a very long time 
but principally because of the Galaxy's gravitational field; yet they did not quantify it further). 

On the other hand, \citet{gardiner96} predicted from their model that the Bridge gas originated 
primarily from the halo of the SMC, not its disk. Under the assumption that the tidal
models are correct, this would mean that despite the steep increase of star
formation rate in the SMC disk, the metallicity of the SMC halo would not have been 
enhanced in the last 2.5 Gyr by any disk material via outflows or else was diluted 
by an external low-metallicity source.
Assuming that the initial mass of the Bridge is $M_{\rm MB}^i$ and metallicity
is $Z_{\rm MB}^i$, and a ``diluting'' gas with  $M_{\rm dil}$ and $Z_{\rm dil}$, 
the  metallicity of the mixed-up gas in the Bridge, $Z_{\rm MB}^f$, can be expressed as: 
\begin{equation}
Z_{\rm MB}^f = \frac{Z_{\rm MB}^i M_{\rm MB}^i + Z_{\rm dil} M_{\rm dil}}{M_{\rm MB}^i + M_{\rm dil}}\,;
\end{equation}
and assuming that $Z_{\rm MB}^i M_{\rm MB}^i \gg Z_{\rm dil} M_{\rm dil}$, we have
\begin{equation}
Z_{\rm MB}^f = Z_{\rm MB}^i \left(1+ \frac{M_{\rm dil}}{M_{\rm MB}^i}\right)^{-1}\,. 
\end{equation}
Therefore, if the metallicity of the tidally disrupted material was present-day SMC like,  
$Z_{\rm MB}^i = 0.25 Z_\odot$, one would need  $M_{\rm dil} \approx 2 M_{\rm MB}^i$ 
(with $Z_{\rm dil} \la 0.01 Z_\odot$) to achieve $Z_{\rm MB}^f \approx 0.09 Z_\odot$. In contrast,  
$Z_{\rm MB}^i M_{\rm MB}^i \ll Z_{\rm dil} M_{\rm dil}$, $Z_{\rm dil}^i = 0.09 Z_\odot$
and $Z_{\rm MB}^i = 0.25 Z_\odot$ would require $M_{\rm dil} \ga 200 M_{\rm MB}^i$ 
to have $Z_{\rm MB}^f = 0.09 Z_\odot$. 
A determination of the ratio the SMC halo to SMC disk particles that were pulled into 
the Bridge in simulations would be valuable. This scenario might be plausible since
dwarf galaxies can have the \hi\ gas occupying radii quite larger than the 
stellar component \citep{salpeter96}. This scenario was also proposed 
by \citet{harris07} to explain the absence of evidence of tidally stripped 
stars in the Bridge. If this envelope was extended enough, it could have reduced 
the impact of feedbacks from massive stars in the SMC at large 
distances from the SMC for $\sim$2--3 Gyr.  

If the very low metallicity material did not come from an extended halo of gas around 
the SMC stellar disk (and assuming the Bridge is young),  the Bridge could
have swept up metal-poor ambient matter over time. For example, 
measurements of the proper motion of the Fornax dwarf spheroidal galaxy suggest that the orbit 
of the Magellanic Clouds was crossed by Fornax some 200 Myr ago \citep{dinescu04},
a time coincidentally similar to the believed formation epoch of the Bridge.
The Fornax has a wide range of metallicity from $-2.0$ to $-0.4$ dex solar, 
and peak near $-0.9$ dex \citep{pont04}. While this is speculative, low-metallicity 
material from the Fornax or other sources may have mixed-up with Bridge gas. 

Bekki \& Chiba (2007a,2007b) have recently produced self-consistent chemodynamical 
models of the LMC-SMC-Galaxy interactions and found
a metallicity distribution that peaks at $-0.8$ dex but could extend down 
to $-0.9$ dex. To the best of our knowledge, this is the only model 
that attempts to explain the lower metallicity of Bridge. In this model, a low metallicity
is reached because the Bridge is mostly formed from the SMC halo gas, where the metallicity
is significantly smaller because of an assumed metallicity gradient in the SMC,
although it is not clear that there is or has been such a gradient since
the present-day metallicity in the SMC Wing \citep{lee05} is similar to the SMC. 

\subsection{Concluding Remarks}\label{sec-per}
Our analysis provides the first metallicity of the Magellanic Bridge gas
and shows for the first time that the Bridge is 
substantially ionized. As our sample of sightlines is small, it is, however, not  
certain if large chemical and/or ionization inhomogeneties exist. 
As we argued above, our three sightlines probe the low \hi\ column density
regions of the Bridge (RA\,$\ga 3^{\rm h}$) at very different locations and 
depths. The present-day metallicity of the Bridge derived from 3 B-type 
stars (including DGIK\,975) also suggests some chemical homogeneity \citep{rolleston99,lee05}. 
We note that two of the Rolleston et al. stars are close to the SMC Wing, but
outside the high \hi\ column density regions. In contrast, \citet{lee05}
found an SMC-like metallicity for 4 B-type stars well inside the 
SMC Wing (and in the high \hi\ column density region of the Wing). 
While the samples are small in each region, it seems very unlikely
that current samples probe each an unrepresentative population. 

The difference of 0.4--0.5 dex in the metallicity between these two adjacent
regions is another puzzle, although we note that \citet{lee05} quoted
errors of $\pm0.3$ dex in their abundances and the difference
may actually be smaller. These authors also used the same
method for the SMC Wing and Bridge stars and found a systematic 
difference of about 0.5 dex in the abundance of these stars. 
This difference may be explained (assuming chemical inhomogeneities 
in each region are small) if the Wing and the low \hi\ column density
regions of the Bridge have a different origin or  the star-formation 
rate and initial-mass function were quite different in each region, 
and/or the Wing was not diluted with extremely-low metallicity gas. In the 
former scenario, the Wing must be younger than the rest of the Bridge, so that it was
formed recently from gas and stars stripped from the SMC that have a present-day SMC
metallicity. This would imply that the Bridge itself must be much older
and that the star-formation was  inefficient over a long period 
of time or only started in the last 200--300 Myr as the results of 
\citet{harris07} suggest. In the second scenario, at the epoch the 
Bridge and the Wing were formed, the metallicity should have been 
around  $\la -1.1$ dex in both regions. However, it would seem very unlikely that 
the metallicity of the Wing has increased by 0.5 dex in the last 200 Myr assuming 
this is the age of these structures. In the SMC where star formation is more
important than in the Wing, about 2--3 Gyr were needed 
to increase the metallicity from $-1.1$ dex to the present-day metallicity \citep{pagel98}.
Therefore the second scenario is not likely if these structures are only 200 Myr old. 
In the third scenario, the Bridge outside the Wing regions could have
been diluted with low metallicity gas, while the Wing was not or at a 
neglible level, possibly because the star-formation was more efficient or the
Wing material comes from deeper regions of  the SMC (where the gas has present-day SMC
metallicity) than the rest of the Bridge gas. 
The first and third scenarios are consistent with the findings of evolved
stars in the Wing and not anywhere else in the Bridge \citep{harris07}. 

Hence despite the fact that current tidal simulations reproduce well
the observed \hi\ structures (i.e. the Bridge, Stream, and Leading Arm), they seem
to be in conflict with the above conclusions regarding the metallicities.  
As already mentioned, the velocities of the LMC and SMC adopted in the models 
of \citet{gardiner96} and subsequent models differ from those derived from the recent
proper motion estimates of the SMC and LMC by  over 100 \km\ \citep{kallivayalil06,kallivayalil06b}. 
It is also unclear how bound the SMC and LMC are and were \citep{kallivayalil06}. 
With those updated proper motions, \citet{besla07} strongly suggest that existing numerical models
of the Clouds may no longer be appropriate, and in particular cannot explain anymore
the Stream and Leading Arm.  As discussed 
by \citet{nidever07}, current simulations may also miss important physics, including 
feedback processes such as outflows from the LMC that may interact with the Bridge gas. 
In particular, it is quite important to understand 
feedback and mixing processes in the SMC disk and halo for the last 3 Gyr 
if the Bridge is only 200 Myr old and its main origin is the halo of the SMC. 

Future theoretical investigations using the updated proper motions of the LMC 
and SMC should not only try to reproduce the \hi\ properties but also accommodate the low metallicity of 
the Bridge, the apparent discrepancy of the metallicity between the Bridge and the
SMC Wing, its ionization structure, and the apparent lack 
of old stars in the Bridge \citep{harris07}. Future models should in 
particular investigage if the Bridge can be stable over a long time (2--3 Gyr) or despite an
important burst of star formation within the SMC 2.5 Gyr ago, the SMC halo gas
could maintain a very low metallicity. If none of these conditions are 
satisfied, an important (depending on the chemical homogeneity in the Bridge)
amount of matter in the Bridge must come from somewhere else. If the Bridge
could survive several perigalactic approaches between the SMC and LMC, it
would be interesting to know if the low density and low metallicity regions of the Bridge 
could have been formed some 2--3 Gyr ago, while the SMC Wing would have been 
produced at a latter time, some 200 Myr ago during the last and closest approach 
between the SMC and LMC. 

From an observational point of view, a larger 
stellar and interstellar sample where the abundances can be estimated in 
the Bridge and the SMC Wing would be extremely valuable. The soon to 
be installed Cosmic Origins Spectrograph (COS)  will be particularly useful 
for the study of the gas-phases in the Bridge since fainter targets 
can be observed with it. In particular, to better characterize the ionization 
structure, future COS and deep H$\alpha$ observations will be invaluable. A better
characterization of the stellar population in the Bridge will also help to 
constrain the origin of the ionization, the star-formation history, and the inital-mass 
function in the Bridge. Since the Bridge is the closest gaseous and stellar tidal remnant, and 
since interactions between galaxies are rather common, future observational and 
theoretical efforts offer the unique opportunity to comprehend the
physical processes and origin of such structures.

\section{Summary}\label{sec-sum}
Using far-UV spectra obtained with {\fuse} and {\em HST}/STIS E140M,
we analyse the physical properties and abundances of the Magellanic Bridge gas 
toward three sightlines that are situated at different locations in the Bridge
and probe various depths. These observations provide access to neutral and 
ionized species at sufficient signal-to-noise and resolution to estimate their 
column densities and kinematics. In Table~\ref{t-prop} we summarize the 
abundances and physical properties of the Magellanic Bridge determined from 
our investigation.  Our main findings are summarized as follows:

\noindent
1. Toward one sigthline, we find that the Magellanic Bridge metallicity is 
$[{\rm Z/H}] = -1.02 \pm 0.07$, and toward another 
one $-1.7 < [{\rm Z/H}] < -0.9$.  To derive these quantities we compare 
the column densities of neutral elements (\oi, \ari, \nni) to that of \hi. 
The metallicity of the gas in the Bridge is in excellent agreement with the average metallicity
determined in B-type stars. There is some evidence that N might be less
deficient than usually observed in gas with $[{\rm Z/H}] \approx -1$ and 
even possibly not deficient with respect to O. 

\noindent
2. The very low present-day metallicity in the Bridge
is similar to the SMC before a burst of star formation that occured
about 2.5 Gyr ago. This may not only be a pure coincidence since
interaction between galaxies are believed to create bursts of star
formation within the interacting galaxies. Yet, it is unclear at this 
time if the Bridge could survive subsequent perigalactic passages of 
the LMC with the SMC. Hence the Bridge could be much younger as currently
predicted by tidal models. In this case, 
it would require a high level of dilution with an unidentified component with 
extremely low metallicity; this component must be dominant in order to reach
the present-day Bridge metallicity.

\noindent
3. We determine that the gas of the Bridge is largely ionized toward our 3 sightlines, 
with only $\sim$20\% of the gas being neutral, implying that the largest fraction of the gas mass 
of the Bridge comes from the ionized gas. Toward two sightlines, we find that there are at least two main 
components in absorption, and the component with the lower velocity is systematically more ionized. 
We argue that possible sources for the ionization are the hot stars within and near the Bridge, 
hot gas (indirectly detected via \ovi\ absorption), and photons leaking from the SMC, LMC, 
and Milky Way. 

\noindent
4. From the analysis of \cii*, we find $n_e < 0.03 \sqrt{T_4}$ cm$^{-3}$ in the
nearly fully ionized gas and $n_e < 0.1 \sqrt{T_4}$ cm$^{-3}$ in the
partially ionized gas. Since the gas is dominantly 
ionized, our analysis suggests  that  the overall density of the Bridge gas is extremely 
low. This is consistent with the absence of detection of H$\alpha$ emission in 
the diffuse gas of the Bridge with current observations. 
Denser and less dense regions must also exist in the 
Bridge on account of the multiphase (cooler and hotter gas) nature of the Bridge.

\acknowledgments

We thank the anonymous referee for constructive comments that improved
and strengthened the content of our manuscript. As this paper 
was submitted,  {\em FUSE}\ abruptely failed permanently.
We want to express our gratitude to the {\em FUSE}\  science, hardware, and mission-operation 
teams for providing high-quality FUV spectroscopic data for the last 8 years. 
Support for this research was provided by NASA through grants 
FUSE-NNX06F42G and FUSE-NNX07AK09G. F.P.K. is grateful to AWE Aldermaston for the award 
of a William Penney Fellowship. This research has made use of the NASA
Astrophysics Data System Abstract Service and the Centre de Donn\'ees de Strasbourg (CDS).

\end{document}